\documentclass{elsart}

\usepackage{graphicx}

\usepackage{amssymb,amsmath}

\newcommand*\patchAmsMathEnvironmentForLineno[1]{%
  \expandafter\let\csname old#1\expandafter\endcsname\csname #1\endcsname
  \expandafter\let\csname oldend#1\expandafter\endcsname\csname end#1\endcsname
  \renewenvironment{#1}%
     {\linenomath\csname old#1\endcsname}%
     {\csname oldend#1\endcsname\endlinenomath}}%
\newcommand*\patchBothAmsMathEnvironmentsForLineno[1]{%
  \patchAmsMathEnvironmentForLineno{#1}%
  \patchAmsMathEnvironmentForLineno{#1*}}%
\AtBeginDocument{%
\patchBothAmsMathEnvironmentsForLineno{equation}%
\patchBothAmsMathEnvironmentsForLineno{align}%
\patchBothAmsMathEnvironmentsForLineno{flalign}%
\patchBothAmsMathEnvironmentsForLineno{alignat}%
\patchBothAmsMathEnvironmentsForLineno{gather}%
\patchBothAmsMathEnvironmentsForLineno{multline}%
}

\usepackage{lineno}

\begin{document}

\begin{frontmatter}



\title{The LCFIVertex package: vertexing, flavour tagging and
vertex charge reconstruction with an ILC vertex detector}


\author[ManchU]{D.~Bailey},
\author[OxU]{E.~Devetak},
\author[BrisU]{M.~Grimes},
\author[RAL]{K.~Harder\corauthref{cor1}},
\ead{Kristian.Harder@stfc.ac.uk}
\author[OxU]{S.~Hillert},
\author[OxU]{D.~Jackson},
\author[RAL]{T.~Pinto Jayawardena},
\author[OxU]{B.~Jeffery},
\author[OxU]{T.~Lastovicka},
\author[BrisU]{C.~Lynch},
\author[EdU]{V.~Martin},
\author[EdU]{R.~Walsh}
\author{\hspace{50em}}
\author{{\bf and the LCFI Collaboration:} \hspace{50em}}
\author[LivpU]{P.~Allport},
\author[OxU]{Y.~Banda},
\author[GlasU]{C.~Buttar},
\author[GlasU]{A.~Cheplakov},
\author[BrisU]{D.~Cussans},
\author[RAL]{C.~Damerell},
\author[NijU]{N.~De Groot},
\author[OxU]{J.~Fopma},
\author[OxU]{B.~Foster},
\author[RAL]{S.~Galagedera},
\author[OxU]{R.~Gao},
\author[RAL]{A.~Gillman},
\author[BrisU]{J.~Goldstein},
\author[LivpU]{T.~Greenshaw},
\author[RAL]{R.~Halsall},
\author[OxU]{B.~Hawes},
\author[LivpU]{K.~Hayrapetyan},
\author[BrisU]{H.~Heath},
\author[OxU]{J.~John},
\author[RAL]{E.~Johnson},
\author[OxU]{N.~Kundu},
\author[GlasU]{A.~Laing},
\author[MontU]{G.~Lastovicka-Medin},
\author[OxU]{W.~Lau},
\author[OxU]{Y.~Li},
\author[RAL]{A.~Lintern},
\author[BrisU]{S.~Mandry},
\author[RAL]{P.~Murray},
\author[RAL]{A.~Nichols},
\author[OxU]{A.~Nomerotski},
\author[BrisU]{R.~Page},
\author[GlasU]{C.~Parkes},
\author[OxU]{C.~Perry},
\author[GlasU]{V.~O'Shea},
\author[LancU]{A.~Sopczak},
\author[RAL]{K.~Stefanov},
\author[EdU]{H.~Tabassam},
\author[RAL]{S.~Thomas},
\author[LivpU]{T.~Tikkanen},
\author[RAL]{R.~Turchetta},
\author[RAL]{M.~Tyndel},
\author[BrisU]{J.~Velthuis},
\author[RAL]{G.~Villani},
\author[NijU]{T.~Wijnen},
\author[LivpU]{T.~Woolliscroft},
\author[RAL]{S.~Worm},
\author[OxU]{S.~Yang},
\author[RAL]{Z.~Zhang}

\corauth[cor1]{corresponding author}
\address[BrisU]{University of Bristol, Tyndall Avenue, Bristol BS8 1TL, UK}
\address[EdU]{The University of Edinburgh, School of Physics and Astronomy,
The King's Buildings, Mayfield Road, Edinburgh EH9 3JZ, UK}
\address[GlasU]{University of Glasgow, Dept of Physics and Astronomy,
Kelvin Building, Glasgow G12 8QQ, UK}
\address[LancU]{Lancaster University, Physics Department, Lancaster LA1 4YB, UK}
\address[LivpU]{University of Liverpool, Oliver Lodge Laboratory,
Liverpool L69 7ZE, UK}
\address[ManchU]{The University of Manchester, School of Physics and Astronomy,
Oxford Road, Manchester, M13 9PL, UK}
\address[MontU]{University of Montenegro, Cetinjski put bb, 81 000 Podgorica,
Montenegro}
\address[NijU]{Radboud University Nijmegen, Experimental High Energy Physics,
P.O.~Box 9010, 6500 GL Nijmegen, The Netherlands}
\address[OxU]{University of Oxford, Particle Physics, Denys Wilkinson Building,
Keble Road, Oxford OX1 3RH, UK}
\address[RAL]{STFC Rutherford Appleton Laboratory,
Harwell Science and Innovation Campus, Didcot OX11 0QX, UK}

\begin{abstract}
The precision measurements envisaged at the International Linear Collider (ILC)
depend on excellent instrumentation and reconstruction software.
The correct identification of heavy flavour jets, placing unprecedented 
requirements on the quality of the vertex detector, will be central for the
ILC programme.
This paper describes the LCFIVertex software, which provides tools for vertex
finding and for identification of the flavour and charge of the leading hadron
in heavy flavour jets.
These tools are essential for the ongoing optimisation of the vertex detector
design for linear colliders such as the ILC.
The paper describes the algorithms implemented in the LCFIVertex package,
as well as the scope of the code and its performance for a typical vertex
detector design.
\end{abstract}

\end{frontmatter}

\section{Introduction}
\label{SectionIntroduction}

\subsection{Purpose and scope of the LCFIVertex package}
\label{SubsectionPurposeAndScope}

The { International Linear Collider} (ILC), colliding electrons and
positrons, is envisaged by the particle physics community to be 
the next high energy accelerator for particle physics research.
Its centre of mass energy is planned to initially range from 
$200\text{--}500\,\mathrm{GeV}$ for physics runs and down to 
$91\,\mathrm{GeV}$ for calibration purposes, with the possibility
of a later upgrade to $1\,\mathrm{TeV}$ \cite{RDR:2007}.
The ILC with its well-known momentum and spin state of the interacting
particles will be complementary to the Large Hadron Collider (LHC),
providing capability of precise measurements of new physics phenomena
and indirect studies of phenomena at energy scales well beyond the
direct energy reach of both the ILC and LHC \cite{Behnke:2001qq}.
Overviews of the physics accessible at the ILC are given in the TESLA 
Technical Design Report \cite{Behnke:2001qq} and in the ILC 
Reference Design Report \cite{RDR:2007}.
The complementarity of the ILC and the LHC has been investigated in
detail by the LHC/ILC Study Group \cite{LHCILC:2006}.

Extraction of new physics accessible at the ILC will rely not only on
high quality of the colliding beams, but also on the use of hermetic,
high-precision detector systems to record the signals of the products of 
the collisions, as well as excellent reconstruction software to analyse
these events.
Since ILC physics is expected to be rich in final states with heavy flavour
jets, it will be important to be able to distinguish $b$ jets, for which
the leading hadron contains a bottom valence quark, from $c$ jets 
containing hadrons with a charm valence quark and jets arising from light
($u$, $d$, $s$) quark hadronisation.
Crucial for this ``flavour tag'' is the high precision measurement of the
tracks of charged particles in the innermost detector system, the vertex
detector, permitting reconstruction of the decay vertices of heavy flavour
hadrons. 
As shown at previous experiments, a sufficiently precise and mechanically
stable detector permits the reconstruction of both the primary vertex at
the point where the particle beams collide, and the full decay chain in 
heavy flavour jets.
For example, in a typical $b$ jet containing a $B$ hadron decay 
$5\,\mathrm{mm}$ away from the interaction point (IP), resulting in a
$D$ hadron that decays, e.g., $3\,\mathrm{mm}$ further away, it is often
possible to reconstruct all three vertices from the tracks in the jet.
Jet flavour identification is aided by observables derived from these
vertices, such as the mass and momentum of the leading hadron that 
decayed.
Further, measurement of the vertex charge permits one to determine if 
the heavy flavour parton is a quark or an antiquark, opening up a range
of measurements that would otherwise be inaccessible.

The LCFIVertex package provides software for vertex finding, flavour
tagging and vertex charge reconstruction.
In the current phase of ILC detector {research and development}, the code is intended for the 
optimisation of the vertex detector design.
Furthermore, it is currently being used for optimisation of the overall
ILC detector concepts, which requires flavour tagging in order to study
the benchmark physics processes chosen to assess the performance of 
different detector designs.

This paper describes the functionality provided by the LCFIVertex package,
as well as the performance achieved for a typical ILC vertex detector
design.
Emphasis is on the algorithms; further technical information can be found 
in the software documentation \cite{LCFIVertexDocumentation}.
The paper is structured as follows: the vertex detector design used for 
the performance examples and the software framework to which the 
LCFIVertex code is interfaced are described in the remainder of this 
section.
Section \ref{SectionTrackSelection} describes the track selection used 
for the different parts of the code; in particular, the algorithms to 
suppress tracks stemming from photon conversions in the tracking volume 
and from the decay of short-lived $\Lambda$ and $K_{S}$ hadrons are 
described.
Section \ref{SectionZVTOP} explains the vertex finding algorithms, 
Section \ref{SectionFlavourTag} the flavour tag and neural net software 
on which the flavour tagging approach relies and Section 
\ref{SectionQuarkCharge} the algorithm for quark charge reconstruction.
Examples of the resulting performance for a typical detector design,
using a sample of $e^{+}e^{-} \rightarrow \gamma/Z \rightarrow q\bar{q}$
with $q = u,\ d,\ s,\ c,\ b$ at a centre-of-mass energy 
$\sqrt{s} = 91.2\,\mathrm{GeV}$, unless otherwise stated, are presented 
in addition to the algorithms.
Section \ref{SectionSummary} gives a summary of the paper.
Software versions and parameter settings used to produce the results in this paper
are described in Appendices A--C.

\subsection{The detector design used for performance evaluation}
\label{SubsectionDetectorGeometry}

The results presented in this paper were obtained with the detector 
model \verb|LDCPrime_02Sc| \cite{LDCPrime02Sc}.
This detector design evolved from the earlier TESLA detector geometry
\cite{Behnke:2001qq} and was implemented in a Geant4-based detector 
simulation by the LDC { (``Large Detector Concept'')} study group.
It relies { mostly} on a pixel-based silicon vertex detector and a time projection
chamber (TPC) to provide charged particle tracking.
{ These detectors are located} inside a solenoid which provides a
magnetic field of $3.5\,\mathrm{T}$.
Also inside the solenoid are fine grained electromagnetic (ECAL) and
hadronic (HCAL) calorimeters.
Additional tracking and calorimetry is foreseen in the forward region,
which is particularly important at ILC energies, given that many of the
relevant physics processes are expected to give rise to multi-jet final
states with at least one jet in the forward direction. The tracking detector layout
of the \verb|LDCPrime_02Sc| detector model is shown in Figure~\ref{FigureDetectorModel}.

{
\begin{figure}[htb]
\begin{center}
\setlength{\unitlength}{\columnwidth}
\begin{picture}(0.95,0.5)
\put(0.05,0){\includegraphics[width=0.9\columnwidth]{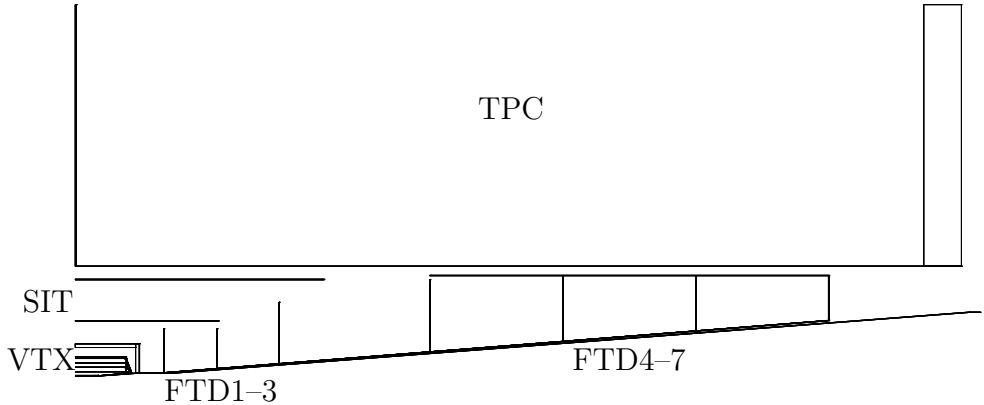}}
\put(0.45,0.26){TPC}
\put(0.0,0.02){VTX}
\put(0.015,0.075){SIT}
\put(0.15,-0.01){FTD1--3}
\put(0.54,0.02){FTD4--7}
\end{picture}
\caption{\textit{Tracking system layout of the LDCPrime\_02Sc detector model
used for the LCFIVertex performance studies presented in this paper. The vertex
detector is shown with cryostat and cables as implemented in the detector simulation.
The outer forward tracking disks FTD4--7 are simulated within a support cylinder. The
TPC endcap is shown on the right. The plot does not cover the entire radial extent of
the TPC.}}
\label{FigureDetectorModel}
\end{center}
\end{figure}
}

The vertex detector geometry in \verb|LDCPrime_02Sc| consists of five
barrel layers of silicon sensors evenly spaced between the inner
radius of $15\,\mathrm{mm}$ and the outer radius of $60\,\mathrm{mm}$.
The length of the active area is $250\,\mathrm{mm}$ for the outer four
layers and $100\,\mathrm{mm}$ for the innermost layer.  The number of
sensor staves per layer ranges between $10$ and $18$.  Sensors are
assumed to be mounted onto a carbon fibre support structure, with the
combined material budget of sensor and support structure corresponding
to $0.1\,\%$ of a radiation length per layer.  As carbon fibre is not
among the materials available in Geant, the support is described as a
$0.134\,\mathrm{mm}$ thick layer of graphite in the simulation.  At
the ends of the barrel staves, the amount of material is larger due to
the electronics and mechanical fixtures needed.  In the barrel region,
the silicon intermediate tracker (SIT), consisting of two layers of
silicon sensors located at radii $160\,\mathrm{mm}$ and
$270\,\mathrm{mm}$, respectively, helps link the track segments
measured in the vertex detector to those provided by the TPC.  In the
detector design considered, the TPC drift region has an inner radius
of $37.1\,\mathrm{cm}$, an outer radius of $180\,\mathrm{cm}$ and a
half-length of $224.8\,\mathrm{cm}$.  In the forward direction,
tracking is complemented by a silicon strip detector, the forward
track detector (FTD), comprised of $7$ disks located at $z$ positions
ranging from $235\,\mathrm{mm}$ to $1997.5\,\mathrm{mm}$ with inner
radii between $23.8\,\mathrm{mm}$ and $162.7\,\mathrm{mm}$ and outer
radii between $140\,\mathrm{mm}$ and $280\,\mathrm{mm}$.  The inner
radius of the silicon-tungsten ECAL is $182.5\,\mathrm{cm}$ and its
outer radius is $201.1\,\mathrm{cm}$.  The ECAL is surrounded by an
iron-scintillator HCAL with an inner radius of $202.5\,\mathrm{cm}$.
Calorimeter cells are squares with sides of length $5\,\mathrm{mm}$ in
the ECAL and $30\,\mathrm{mm}$ in the HCAL.  In the forward region, an
ECAL endcap extends from $z = 245\,\mathrm{cm}$ to $z =
264\,\mathrm{cm}$, followed by an HCAL from $z = 294\,\mathrm{cm}$.

\subsection{LCIO data format and the MarlinReco software framework}
\label{SubsectionLCIOMarlinReco}

The results presented in this paper were obtained with input from
the Pythia event generator \cite{Sjostrand:2006za}.
The detector response was simulated using the Geant4-based
\cite{GEANT4:2003:2006} program MOKKA \cite{MoradeFreitas:2004sq} 
and the detector model \verb|LDCPrime_02Sc|, described above.

The Linear Collider I/O (LCIO) persistency framework \cite{Gaede:2003ip}
permits storage of results between the stages of detector simulation,
event reconstruction and physics analysis, and the exchange of these
results between the different software frameworks used for ILC 
detector optimisation.
One of the main roles of LCIO is to provide a common data model,
in which an \verb|Event| entity holds collections of objects relevant
to the different stages of reconstruction, as well as some of the 
Monte Carlo (MC) information from the event generator.

Reconstruction was performed using tools from the MarlinReco event
reconstruction package \cite{Wendt:2007iw}, based on the particle flow
concept and implemented within the modular C++ application framework
Marlin \cite{Gaede:2006}.
The modules are called ``processors'' in this framework and can be
configured by steering files in xml format.

Specific MarlinReco {processors} were used to simulate the digitization
in the different detector subsystems.
In the case of the silicon-based detectors, the physical processes
occurring in the silicon sensors, such as the drift of the electrons
and holes that give rise to the signals, are described using simple
parameterisations.
Track finding, including pattern recognition, and track fitting was
then performed using the \verb|FullLDCTracking| processor
\cite{Raspereza:2007,Raspereza:2008}.

Calorimeter clusters and tracks are matched using a particle flow
algorithm (PFA).
For this purpose, the PandoraPFA code \cite{Thomson:2006}
was used, resulting in a collection of \verb|ReconstructedParticles|.
These are the entities in LCIO on which analyses are usually based, 
and which are used, for example, as the input to jet finding 
algorithms.
The LCFIVertex code is run on jets, and hence for studies in the 
context of PFA-oriented detectors, like the one described here,
requires the PFA code to be run before the jet finding step.
Calorimeter clusters are required by the PFA algorithm as part of
the input, and therefore also need to be reconstructed before this
algorithm can be run.
Via the jet energy, calculated in the PFA code, calorimeter clusters
are indirectly used for the flavour tag.
Otherwise, the LCFIVertex code is independent of the calorimeter 
information and only accesses the tracks in the input jets.
It was checked that when running a different particle flow
algorithm available in MarlinReco, Wolf \cite{Raspereza:2006}, 
flavour tagging performance did not change.
The Durham $k_{T}$-cluster jet finding algorithm, as implemented in
the \verb|SatoruJetFinder| package in MarlinReco \cite{Yamashita:1998},
was then run, forcing each event into a two-jet topology.
Tracks from photon conversions, $K_{S}$ and $\Lambda$ decays were
identified by a conversion tagger that forms part of LCFIVertex, 
see Section \ref{SubsectionConversionTagger}.
These tracks were excluded from the input passed to the LCFIVertex
processors.

A simple processor to determine the event vertex or interaction point
(IP) from an iterative fit to a subset of all tracks in an event
was run.
This IP-fit processor and the conversion tagger are the only 
parts of the LCFIVertex package that are not jet-based.
The resulting event vertex is used both by the vertex finding
algorithms described below and for the flavour tag to determine a 
track's point of closest approach to the IP.

\section{Track Selection}
\label{SectionTrackSelection} 

The default flavour tag procedure implemented in the LCFIVertex
package, as described in more detail in Section \ref{SectionFlavourTag},
uses secondary vertex information whenever available.
Different neural networks with separate sets of input variables
are used depending on whether secondary vertices are found or not.
In particular, for distinguishing $c$ jets from $uds$ jets, a 
useful criterion is that $uds$ jets do not contain vertices 
stemming from the decays of heavy flavour hadrons.
It must therefore be ensured that the vertex finder creates as few
fake vertices from wrong track combinations as possible.
This can partly be achieved by appropriate track selection, as 
discussed in Section \ref{SubsectionConversionTagger}.
Other potential sources of unwanted vertices for the purpose of
flavour tagging are photon conversions in the detector material
and the decay of $K_{S}$ and $\Lambda$ particles, which can resemble
a heavy flavour decay vertex.
As these effects have clear signatures, such tracks can easily be
identified and suppressed by the track selection.

\subsection{Identification of tracks from photon conversions, $K_{S}$
and $\Lambda$ decays}
\label{SubsectionConversionTagger}

{A dedicated Marlin processor is used to identify tracks from $K_{S}$ and $\Lambda$ 
decays and from photon conversions.
All two-track combinations are considered as candidates and}
are required to pass the following
criteria to be identified as conversions or $K_{S}$/$\Lambda$
decays:
\begin{itemize}
\item{The constituent tracks must have opposite charge sign.}
\item{The distance of closest approach between the two track 
helices must not exceed $1\,\mathrm{mm}$.}
\item{The distance between the point of closest approach and
the IP must be larger than $1\,\mathrm{mm}$ in order to 
reduce the risk of fake tags consisting of combinations of
primary vertex tracks.}
\item{The invariant mass of the two track combination has to
be compatible with that of the photon, the $K_{S}$ or the
$\Lambda$.}
\end{itemize}
To check the mass compatibility, the rest mass of the combination is
calculated using three mass hypotheses, choosing the masses of the 
decay products accordingly:
both tracks are assumed to be electrons in the case of conversions,
or pions for the $K_{S}$ hypothesis, and for the 
$\Lambda$ hypothesis, the track with larger momentum is assigned the
proton mass and the other the pion mass.
The resulting rest mass of the combination is considered to be
compatible with the hypothesis if it differs from the PDG value by 
not more than $5\,\mathrm{MeV}$ for conversions and Kaons and by not
more than $2\,\mathrm{MeV}$ for Lambdas. {Table~\ref{TableConvTagPerformance}
lists performance figures for each particle type separately.}
{
\begin{table}[htb]
\begin{center}
\begin{tabular}{||c||c|c||}
\hline
particle  & efficiency & purity \\
\hline
photon    &    24.6\%  & 96.3\%  \\ \hline
$K_S$     &    72.2\%  & 67.4\%  \\ \hline
$\Lambda$ &    69.8\%  & 62.3\%  \\ \hline
\end{tabular}
\vspace*{2ex}
\caption{\textit{Performance of the conversion, $K_S$ and $\Lambda$
    identification.  The efficiency is normalised to particles where
    all conversion respectively decay tracks were reconstructed in the
    detector, i.e.~to particles that could potentially lead to
    reconstruction of additional vertices. The purity is defined as
    the fraction of reconstructed conversion/$K_S$/$\Lambda$
    candidates with a correct combination of tracks. Selection cuts
    for each particle type were chosen to optimise flavour tag
    performance, not to optimise standalone performance of the
    conversion/$K_S$/$\Lambda$ tagger.}}
\label{TableConvTagPerformance}
\end{center}
\end{table}
}
All objects that are not identified as stemming from the above sources
are passed on to the track selection processor preceding the subsequent
steps such as vertex finding and flavour tagging.

In addition to this core functionality, the conversion tagger
can also be run in one of two ``cheater'' modes, which use MC information
to identify conversions and $K_{S}$ or $\Lambda$ decays and can be used
to assess the performance of the realistic reconstruction.
The two cheater modes differ in the way they treat the case that only one
of the tracks from a conversion or $K_{S}$/$\Lambda$ decay has been
reconstructed, the other track not being within the detector acceptance or
being lost due to pattern recognition inefficiencies.
The more moderate cheater mode then does not flag the track that could be
reconstructed as resulting from a conversion or $K_{S}$/$\Lambda$ decay,
since no realistic algorithm could possibly identify such tracks, whereas
the more aggressive cheater mode flags all tracks that stem from these
sources.

\begin{figure}[htb]
\begin{center}
\includegraphics[width=0.95\columnwidth]{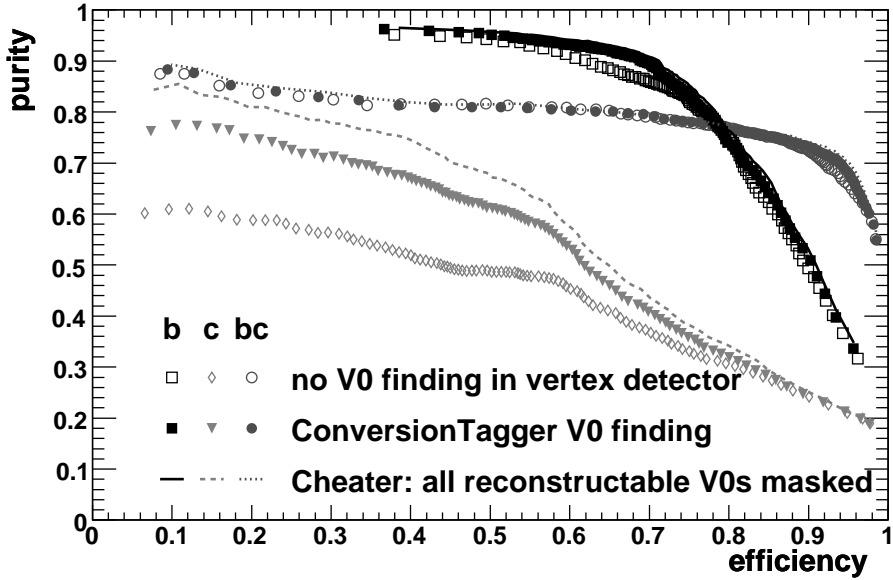}
\caption{\textit{Comparison of tagging with and without conversion
    tagging and performance obtained when using MC information at
    $\sqrt{s} = 91.2\,\mathrm{GeV}$.  {The horizontal axis shows the
    efficiency with which true b (c) jets are tagged, and the vertical
    axis shows the fraction of true b (c) jets in the sample of tagged
    jets. The performance is evaluated separately for b tagging, c tagging
    and c tagging within a sample of b and c jets only (bc).}}}
\label{FigureConversionTaggerPerformance}
\end{center}
\end{figure}
Figure~\ref{FigureConversionTaggerPerformance} shows the flavour tagging 
performance in terms of purity vs.~efficiency for the three tags provided 
(see Section \ref{SubsectionNeuralNets} for details) when the conversion
tagger  is run as part of the reconstruction chain (full 
symbols).
Performance without any $K_{S}$/$\Lambda$ identification and with only
the conversions that PandoraPFA finds removed, i.e.~no conversion
finding in the vertex detector, is plotted for comparison (open symbols).
Also shown is the performance obtained from the moderate cheater mode
(lines).

\subsection{Track selection for IP fit, ZVTOP and flavour tag}
\label{SubsectionTrackSelection}

\begin{table}[htb]
\begin{center}
\begin{tabular}{||c||c|c|c||}
\hline
parameter & IP fit & vertexing (ZVRES) & flavour tag \\
\hline 
$\chi^{2}/\mathrm{ndf}$ of track fit     &   5 & 4     & --- \\ \hline
$R$-$\phi$ impact parameter $d_{0}$ (mm) &  20 & 2     & 20  \\ \hline
$d_{0}$ uncertainty (mm)                 & --- & 0.007 & --- \\ \hline
$z$ impact parameter $z_{0}$ (mm)        &  20 & 5     & 20  \\ \hline
$z_{0}$ uncertainty (mm)                 & --- & 0.025 & --- \\ \hline
track $p_{T}$ ($\mathrm{GeV}$)           & 0.1 & 0.2   & 0.1 \\ 
\hline
\end{tabular}
\vspace*{2ex}
\caption{\textit{Track selection used for IP fit, vertexing and calculation
of the flavour tag inputs.}}
\label{TableTrackSelection}
\end{center}
\end{table}
The track selection can be tuned separately for the IP fit, the ZVTOP
vertex finder and the calculation of the flavour tag inputs.
An overview of the track selection cuts for each of these tasks is given 
in Table \ref{TableTrackSelection}.
In addition to the cuts listed in the table, requirements are
implemented on the number of hits in the tracking subdetectors as follows:
if there are at least $20$ hits in the TPC or at least three hits in the
FTD, {no} hit is required in the vertex detector.
If there are fewer hits in the TPC or FTD, at least three vertex detector
hits are needed.

\section{The ZVTOP vertex finding algorithms}
\label{SectionZVTOP}

{The LCFIVertex package contains a complete re-implementation of the
topological vertex finder ZVTOP originally developed at the SLD
experiment \cite{Jackson:1996sy}. A modified version of the original
Fortran code was used in previous Linear Collider studies, see
e.g.~Refs.~\cite{RDR:2007,Behnke:2001qq}. For the most part, the
vertex finding algorithm in LCFIVertex, as described below in this
section, corresponds to the original SLD version. Minor improvements
of the LCFIVertex algorithm with respect to SLD include a Kalman
vertex fit and adjustments to allow use of ZVTOP in events at
centre-of-mass energies above the $Z$ resonance.}

\subsection{The ZVRES algorithm}
\label{SubsectionZVRES}

{The ZVTOP vertex finder provides two complementary 
algorithms} which use topological information to identify track combinations 
that are likely to have their origin at a common vertex.
The first of these, the ZVRES algorithm, can be used to find multi-pronged
secondary vertices with an arbitrary geometrical distribution and hence is
most generally applicable provided the detector system has a sufficiently 
high spatial resolution.
The object-oriented implementation of the code that forms part of the
LCFIVertex package is described in detail elsewhere \cite{Jeffery}.
Compared to the original implementation, the new code provides several
improvements and adjustments to a collider environment with a variable
centre-of-mass energy.
The remainder of this section outlines the vertexing algorithms and 
describes these modifications.

A central idea of the ZVRES algorithm is to describe each track $i$ by
a probability density function $f_{i}(\vec{r})$ in three-dimensional
space and to use these to define a vertex function $V(\vec{r})$ that
yields higher values in the vicinity of true vertex locations and lower
values elsewhere, as well as providing a criterion for when two vertex
candidates are resolved from each other.

The track functions have a Gaussian profile in the plane normal to the
trajectory.
With $\vec{p}$ the point of closest approach of track $i$ to space
point $\vec{r}$, the track function $f_{i}(\vec{r})$ is defined as:
\[ f_{i}(\vec{r}) = \exp 
                     \left\{
		       - \frac{1}{2} 
		          \left(  \vec{r} - \vec{p}
			  \right)
			  \mathbb{V}_{i}^{-1}
			  \left(  \vec{r} - \vec{p}
			  \right)^{T}
                     \right\} \ \ \ ,
\]
where $\mathbb{V}_{i}$ is the position covariance matrix of the track at 
$\vec{p}$.

In its most basic form, the vertex function is defined as
\[ V(\vec{r}) = 
         \sum_{i=1}^{N} f_{i}(\vec{r}) -
	 \frac{\sum_{i=1}^{N} f_{i}^{2}(\vec{r})}{\sum_{i=1}^{N} f_{i}(\vec{r})}
\]
with the second term ensuring that $V(\vec{r})$ approaches zero in spatial
regions in which only one track contributes significantly to the first
term and where hence no vertex should be found.

Optionally, further knowledge on where vertices are more likely to be
found can be used to weight the vertex function, thereby suppressing fake
vertices and increasing the purity of the vertices found (i.e.~the fraction
of correctly assigned tracks).
Knowledge of the IP position can be used to suppress fake vertices from
tracks passing close by each other in the vicinity of the IP.
This is accomplished by representing the IP by a contribution
\[ f_{0}(\vec{r}) = \exp 
                     \left\{
		       - \frac{1}{2} 
		          \left(  \vec{r} - \vec{p}
			  \right)
			  \mathbb{V}_{IP}^{-1}
			  \left(  \vec{r} - \vec{p}
			  \right)^{T}
                     \right\} \ \ \ ,
\]
where $\vec{p}$ is the position of the IP and $\mathbb{V}_{IP}$ the
covariance matrix describing the accuracy with which this position 
is known.
This new term contributes to the vertex function in the same way as the
Gaussian probability functions representing the tracks, hence the
vertex function is redefined as
\[ V(\vec{r}) = 
         w_{IP} f_{0}(\vec{r}) + \sum_{i=1}^{N} f_{i}(\vec{r}) -
	 \frac{w_{IP}^{2} f_{0}^{2}(\vec{r}) + \sum_{i=1}^{N} f_{i}^{2}(\vec{r})}
	      {w_{IP} f_{0}(\vec{r}) + \sum_{i=1}^{N} f_{i}(\vec{r})}
\]
This definition ensures that space points close to the IP are less likely
to be resolved from each other and that tracks that could otherwise give
rise to fake vertices are more likely to be assigned to the primary
vertex.
In the default configuration of the code, the IP contribution is given 
a weight of $w_{IP} = 1$.

\begin{figure}[htb]
\begin{center}
\includegraphics[width=0.5\columnwidth,clip=true]{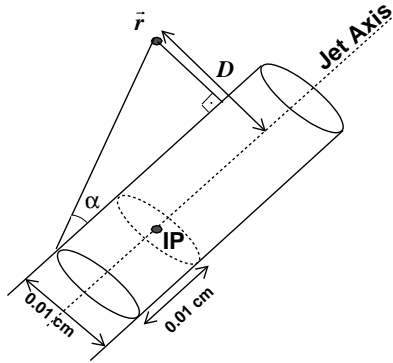}
\caption{\textit{Schematic diagram showing the definition of angle $\alpha$.}}
\label{FigureKalpha}
\end{center}
\end{figure}
Similarly, it is kinematically favoured for heavy hadrons to decay 
close to the jet axis rather than at large angle from it, which is
taken into account by weighting the vertex function outside a cylinder
of radius $50\,\mu\mathrm{m}$ by the factor $\exp^{-K_{\alpha}\alpha^{2}}$.
In this expression, the angle $\alpha$ is defined as shown in 
Fig.~\ref{FigureKalpha} and $K_{\alpha} = k E_{\mathrm{Jet}}$ with $k$ 
being an LCFIVertex parameter which the user can set and $E_{\mathrm{Jet}}$ the 
jet energy.
The jet energy dependent definition of $K_{\alpha}$ implemented in
LCFIVertex takes into account that jets of higher energy are more
collimated.

In addition to indicating likely vertex positions, the other main
use of the vertex function is to provide a key criterion for 
merging candidate vertices in the process of vertex finding: 
space points $\vec{r_{1}}$ and $\vec{r_{2}}$ are defined to be
resolved from each other if along the straight line connecting 
these points the vertex function falls below a given fraction
$R_{0}$ of the lower of the values $V(\vec{r}_{1})$ and
$V(\vec{r}_{2})$.
Vertices that are not resolved according to this criterion will
be referred to as unresolved from each other.

The main challenge of many vertex finders used prior to ZVTOP
is the large number of track combinations that need to be 
considered to determine whether they form a good vertex.
In contrast, the ZVRES algorithm uses a bottom-up approach,
starting out from all possible two-track combinations and
using the vertex function as well as the fit $\chi^{2}$ to
decide which candidates to keep and to merge.

In the initial step, all two-track and optionally all
track-IP combinations are fitted with a Kalman vertex fit.
Of these two-object fits, those with $\chi^{2}$ lower than a
threshold $\chi_{0}^{2}$ and vertex function at the fitted 
vertex position $\vec{r}_{\mathrm{Vert}}$ above a threshold 
$V_{0}$ are retained.
Before beginning the merging, the number of remaining 
two-object fits to be further considered is reduced as 
follows: 
for each track in turn, all the two-object combinations that
contain the track are considered.
The track is removed from all vertices with 
$V(\vec{r}_{\mathrm{Vert}})$ below $10\,\%$ of 
$V_{\mathrm{max}}(\vec{r}_{\mathrm{Vert}})$, the maximum
vertex function value obtained from the fits for the track
under consideration.
The track's two-object fits are sorted with respect to their
vertex function $V(\vec{r}_{\mathrm{Vert}})$.
All vertex candidates for which both objects have been
removed at this stage are discarded.

The remaining candidate vertices are then merged making
further use of the resolvability criterion:
starting out from the candidate vertex with the highest
vertex function in the sorted list found in the previous 
step, a set of unresolved vertices is found by iteratively
adding other candidates that are unresolved from any of 
the vertices in the set.
The resulting set of unresolved candidate vertices is then
merged to form a new candidate.
From the remaining vertices, the next seed in the sorted list
is picked and the process continues until all original 
candidate vertices have either been absorbed or considered
as seeds for an unresolved set.
Note that for the merging phase, the vertex function is not
evaluated at the position $\vec{r}_{\mathrm{Vert}}$ of the 
original two-object candidate vertex, but at the closest
local maximum $\vec{r}_{\mathrm{MAX}}$ of the vertex function,
in order to improve the suppression of fake vertices.

Following the merging phase, tracks with a high $\chi^{2}$
contribution are removed from the resulting candidate vertices:
iteratively, the track with the highest $\chi^{2}$ contribution
is removed if its $\chi^{2}$ is above a threshold 
$\chi_{\mathrm{TRIM}}^{2}$.
The vertex is refitted and this step is repeated until the 
highest $\chi^{2}$ track passes the cut.
Candidates that, after this procedure, are no longer associated
with either at least two tracks or with the combination of the
IP object and $0$, $1$ or more tracks, are discarded.

Remaining ambiguities in the association of tracks to vertices
are resolved by keeping each track only in the vertex candidate
with highest $V(\vec{r}_{\mathrm{MAX}})$ and removing it from
all others.
The resulting track combinations are fitted to yield the final
vertices, which are sorted with respect to their distance from
the IP.

\subsection{The ZVKIN algorithm}
\label{SubsectionZVKIN}

The ZVTOP vertex finder also provides an algorithm, ZVKIN, to address
a particular category of jets for which the ZVRES 
algorithm fails, namely $b$ jets with two subsequent
one-prong decays (i.e.~decays in which only one charged track
is detected), see e.g.~Ref.~\cite{Thom:2002} for a description of
the algorithm at SLD, and Ref.~\cite{Jeffery} for further details
of the C++ implementation.
For cases with one multi-prong vertex and one single-prong vertex
a recovery algorithm can identify the decay track and subsequently
add it to the decay chain (see Section 
\ref{SubsectionFlavourTagInputs} for details).
However, if only one charged track can be detected from both the
$B$ vertex and the $D$ vertex, a different approach is needed.
The solution chosen in the ZVKIN algorithm is to use additional
kinematic information by approximating the direction of flight of
the $B$ hadron and forming a ``ghost track'' from it.
This ghost track, described as a straight line with an appropriate
circular error ellipse of constant width along the track, is added 
to the set of tracks from which vertices are found.
The algorithm consists of two main stages: first the ghost track
direction and width are found using an iterative $\chi^{2}$
minimisation approach.
In the second stage this ghost track is used to constrain the
secondary vertex finding.

The jet axis is chosen to initialise the ghost track direction.
Fixed at the position of the IP, the ghost track $G$ is swivelled
in both $\theta$ and $\phi$ directions, until the value
\[
\chi_{S1} = \left\{
\begin{array}{ccc}
\sum_{i}{\chi_{i}^{2}}                                  & , \mathrm{if} & L_{i} \geq 0 \\
\sum_{i}{\left( 2 \chi_{0i}^{2} - \chi_{i}^{2} \right)} & , \mathrm{if} & L_{i} <    0 \\
\end{array}
\right .
\]
is minimal, where $\chi_{i}^{2}$ is the $\chi^{2}$ of the vertex fit
found using track $i$ and $G$ and $\chi_{0i}^{2} = \chi_{i}^{2}(L_{i} = 0)$,
and $i$ runs over all input tracks.
The value $L_{i}$ is the distance from the IP to the projection of the
vertex onto $G$, with the sign chosen to be positive if the vertex is in 
the hemisphere defined by the direction of the jet axis.
Minimisation continues until changes in both $\theta$ and $\phi$ by
$0.1\,\mathrm{mrad}$ do not yield further improvement.
If the initial direction of $G$ is relatively far from the true line of
flight of the $B$ hadron, then some, or possibly all, of the secondary
tracks will yield a vertex with $G$ that has a negative value of $L_{i}$.
The first minimisation stage is designed so that the contribution to
$\chi_{S1}$ from tracks with $L_{i} < 0$ will tend to push the ghost
track $G$ towards the $B$ flight path in the minimisation process.
The $2 \chi_{0i}^{2}$ term ensures that the $\chi_{S1}$ changes in 
a continuous manner at $L_{i} = 0$.

At this point, the width of the ghost track is calculated such that the
largest value of $\chi_{i}^{2}$ for the vertex fit of any of the tracks
with $G$ is equal to 1.
If it is above the user-settable minimum width $\delta_{\mathrm{G,min}}$
required for $G$, it is used in the next step; otherwise the minimum
width is used.
With the adjusted ghost track width, the ghost track direction is further
optimised, using the same algorithm as before, but this time minimising
the quantity
\[
\chi_{S2} = \left\{
\begin{array}{ccc}
\sum_{i}{\chi_{i}^{2}}  & , \mathrm{if} & L_{i} \geq 0 \\
\sum_{i}{\chi_{0i}^{2}} & , \mathrm{if} & L_{i} <    0\ \ . \\
\end{array}
\right .
\]
The width is then adjusted as before.
This choice ensures that the corresponding vertex probability has
an approximately flat distribution between $0$ and $1$, while fake
vertices yield values close to 0.
The vertex probability is calculated from the $\chi^{2}$ value of 
the vertex fit following the standard algorithm described in
\cite{NumRec:1992}.

The resulting optimised ghost track is then used to find vertices
as follows: from each track in the jet, a vertex is formed with the
ghost track.
The IP is added to this initial set of vertices.
For all possible pairs of vertices from this set, the vertex 
probability is calculated, omitting the ghost track from the 
combinations that contain the IP.
The pair that maximises this probability is merged and replaces the
original vertices in the list.
Combinations with vertex probability below a threshold value
$P_{\mathrm{V,min}}$ are not accepted, while the original vertices
which formed the combination are separately retained.
The merging process continues until no further combination yields
a vertex probability above $P_{\mathrm{V,min}}$ or until there is 
only one vertex left.
Finally, the ghost track is removed from all vertices without 
refitting them, with the possibility that secondary vertices contain
only one track.

\subsection{Performance of vertex finding for a typical ILC vertex
detector design}
\label{SubsectionZvresPerformance}

{All results presented in this section were obtained with the ZVRES
algorithm, described in Section \ref{SubsectionZVRES}, which applies
to a broader class of jets than ZVKIN.}
\begin{figure}[htb]
\begin{center}
\begin{tabular}{c}
\includegraphics[width=1.0\columnwidth]{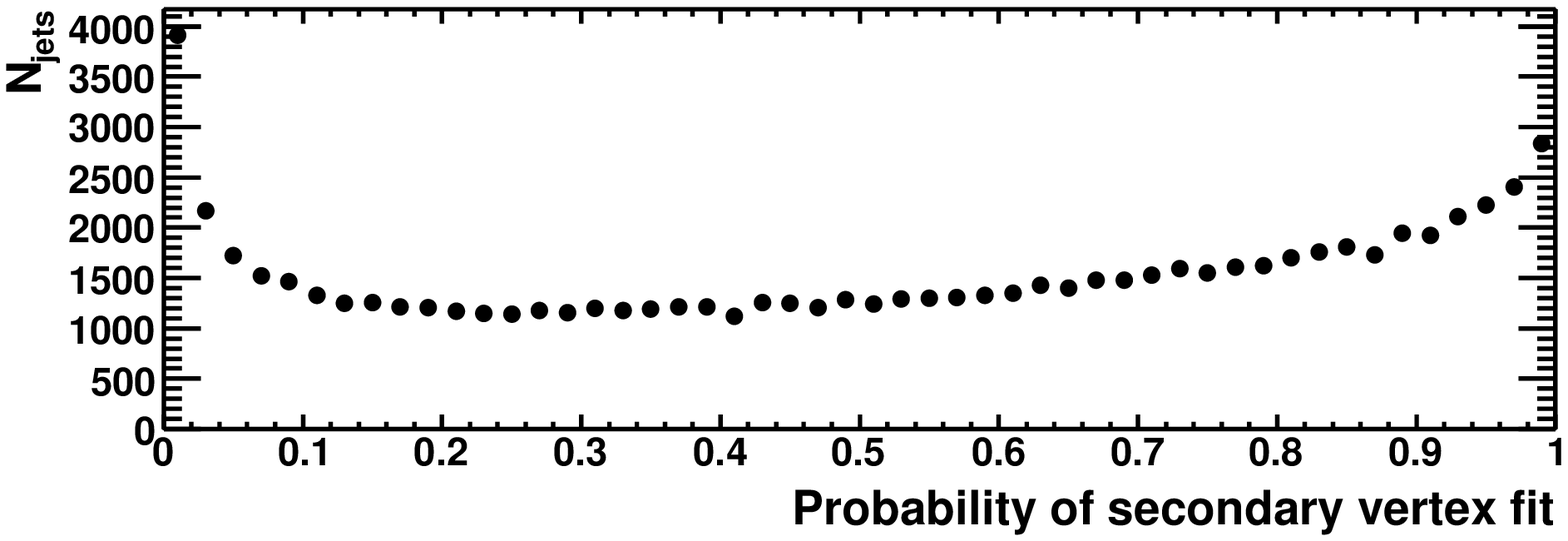}\\[-12ex]
\hspace*{25em}(a) \\[7ex]
\includegraphics[width=1.0\columnwidth]{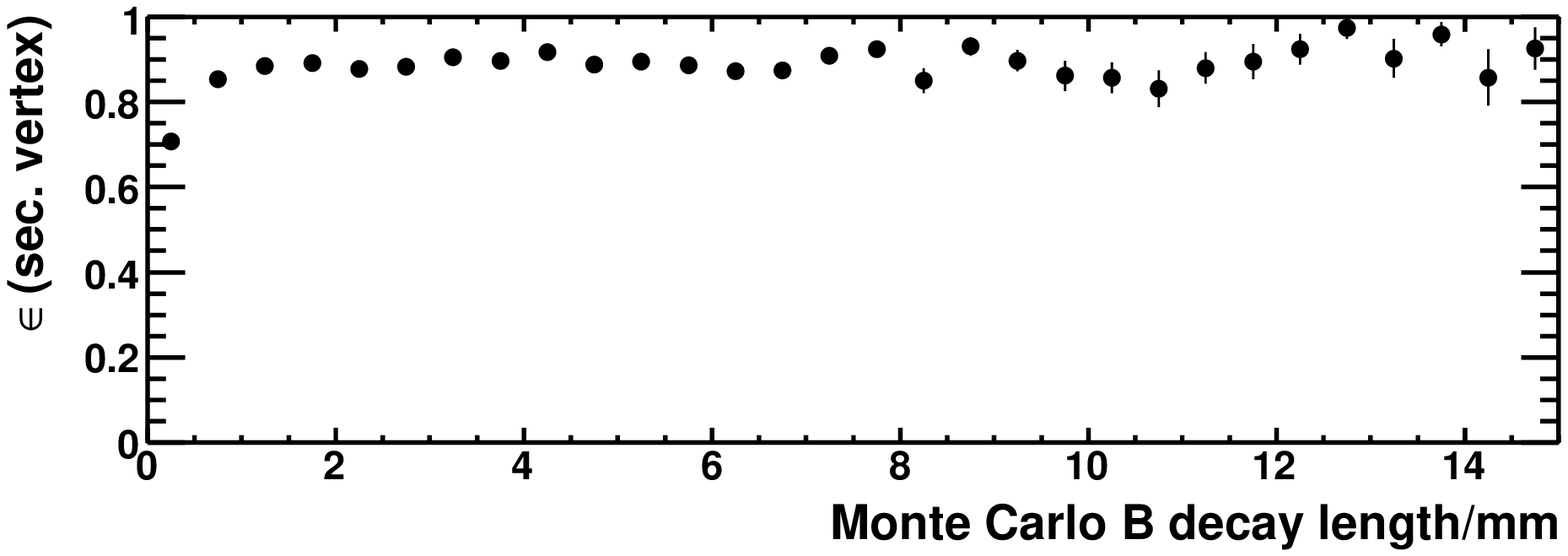}\\[-12ex]
\hspace*{25em}(b) \\[7ex]
\end{tabular}
\caption{\textit{(a) Probability corresponding to the $\chi^{2}$ value calculated from
the vertex fit and (b) efficiency for finding a secondary vertex, for a pure sample of
$b$ jets.}}
\label{FigureZVTOPProbAndEff}
\end{center}
\end{figure}
A pure sample of jets from the process 
$e^{+}e^{-} \rightarrow \gamma/Z \rightarrow b\bar{b}$
at $\sqrt{s} = 91.2\,\mathrm{GeV}$ was used, permitting a direct
comparison with earlier studies at that energy.
Figure~\ref{FigureZVTOPProbAndEff} (a) shows {the $\chi^{2}$ probability 
of the secondary vertex fit.}
This variable cross-checks the vertex fitter, a reasonably flat 
distribution, as seen in the figure, indicating that the fitter 
performs as expected.
For a study of the dependence of vertex finding on the decay length
of the $B$ hadron, a sub-sample was selected, consisting only of jets
in which a $B^{\pm}$ hadron decayed to a charged $D$ hadron, either
directly or via a short-lived $D^{*}$ resonance (with the $D^{*}$
lifetime being so short that it does not travel any measurable distance
in the detector, but decays essentially at its point of origin).
The decay lengths of the $B$ and the $D$ hadron which were chosen 
in the MC simulation were determined for each jet and compared to the
values reconstructed by the ZVRES algorithm.
For this comparison the efficiency of finding a secondary vertex 
within a given decay length interval was plotted as a function of the
true MC $B$ hadron decay length, as shown in 
Fig.~\ref{FigureZVTOPProbAndEff} (b).

\begin{figure}[htb]
\begin{center}
\begin{tabular}{cc}
\includegraphics[width=0.50\columnwidth]{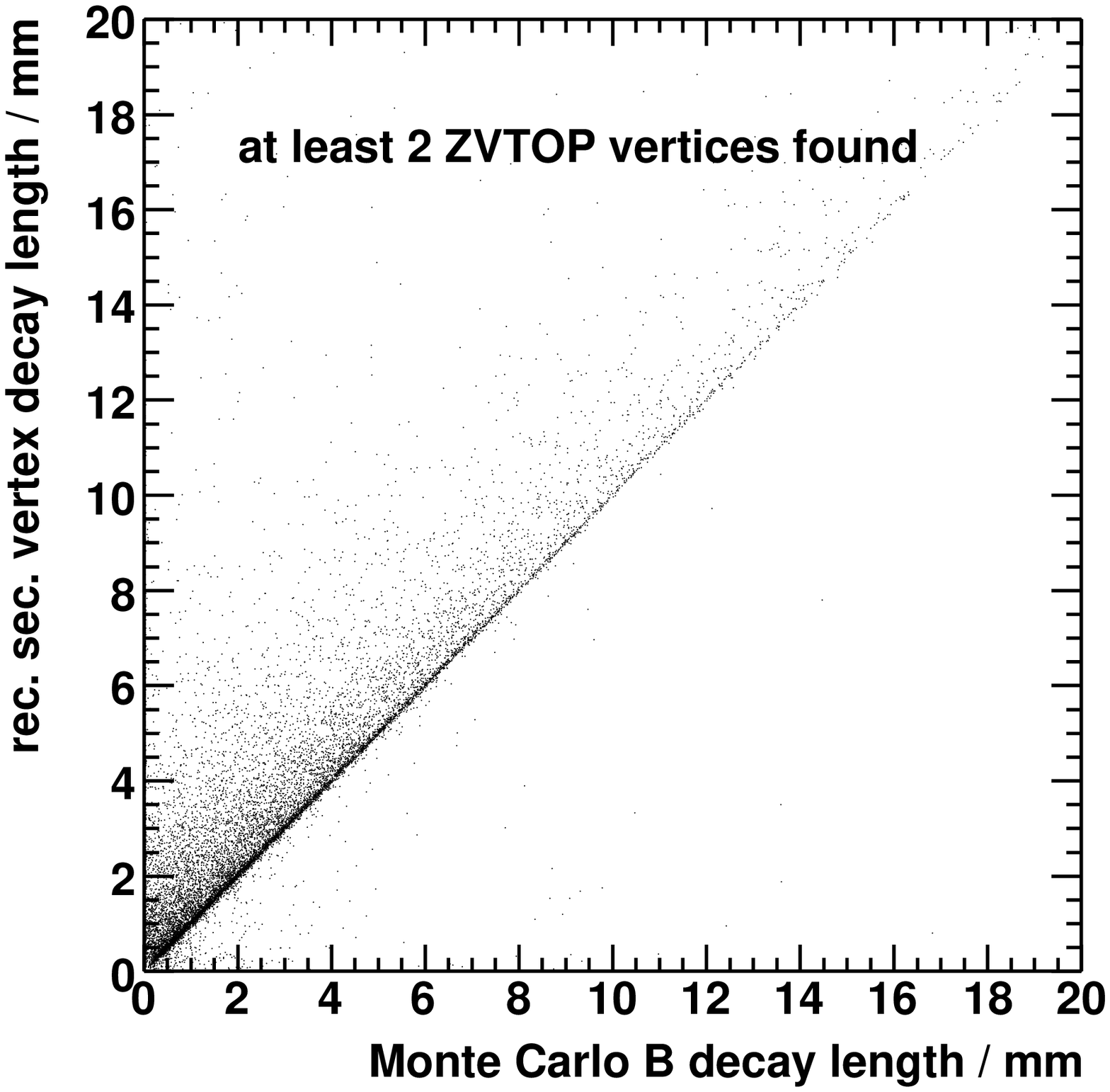} &
\includegraphics[width=0.50\columnwidth]{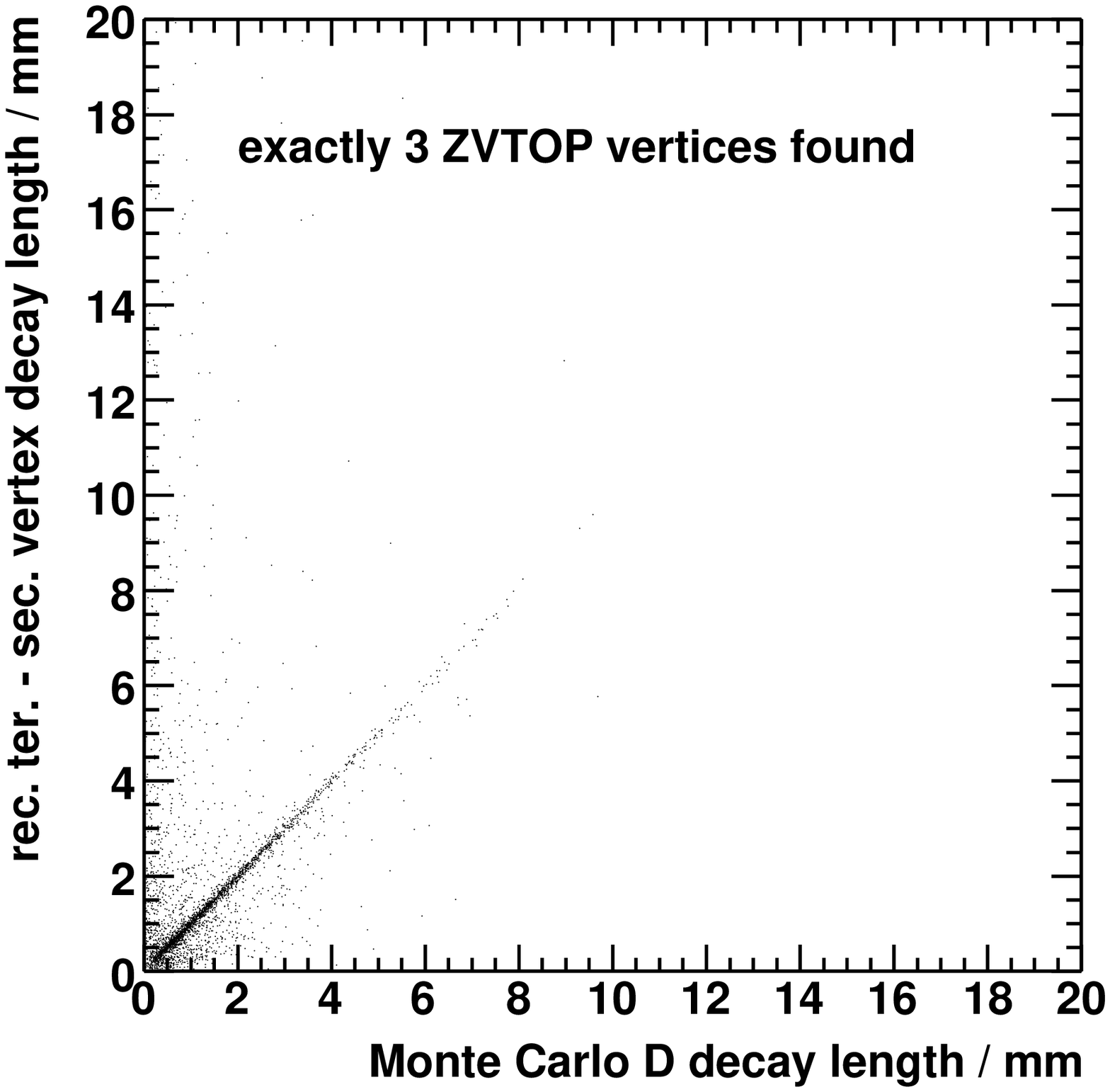}\\[-12ex]
\hspace*{12em}(a) & \hspace*{12em}(b) \\[7ex]
\end{tabular}
\caption{\textit{Reconstructed decay length vs.~MC decay length for
(a) $B$ hadron decays in $b$ jets in which at least two vertices have been found
and (b) $D$ hadron decays in $b$ jets with three ZVTOP vertices.}}
\label{FigureReconstructedVsTrueDecayLength}
\end{center}
\end{figure}
Using the same sub-sample of jets as in Fig.~\ref{FigureZVTOPProbAndEff} 
(b), the reconstructed decay lengths of the $B$ and $D$ hadron were 
compared to the corresponding MC decay lengths on a jet-by-jet basis.
In Fig.~\ref{FigureReconstructedVsTrueDecayLength} (a), the $B$ decay 
length comparison is shown for jets in which at least two vertices 
(the primary, or IP vertex, plus at least one secondary vertex) were 
found.
This class of jets includes cases in which the $D$ hadron decayed so 
close to the $B$ decay vertex that it could not be resolved from it,
and for which therefore the reconstructed decay length is shifted to 
larger values compared to the MC $B$ decay length.
Short $B$ decay lengths, not permitting the $B$ decay vertex to be 
resolved from the primary vertex, also result in deviation of the 
reconstructed secondary vertex decay length from the MC truth value.
In Fig.~\ref{FigureReconstructedVsTrueDecayLength} (b), the decay length 
of the $D$ hadron decay is compared to the MC value for jets in which 
exactly three vertices were reconstructed by ZVRES.
As expected for this category of jets, the correlation between 
reconstructed and MC values is better.

\begin{table}[htb]
\begin{center}
\begin{tabular}{||cc|ccc|cccc||}
\hline
\multicolumn{2}{||c|}{Monte Carlo} &
\multicolumn{7}{c||}{Reconstructed track-vertex association} \\
\multicolumn{2}{||c|}{track origin} &
\multicolumn{3}{c|}{Two vertex case} &
\multicolumn{4}{c||}{Three vertex case} \\ 
\hline
 & & $\ \ $pri$\ \ $ & sec & iso & $\ \ $pri$\ \ $ & sec & ter & iso \\
\hline
$91.2\,\mathrm{GeV}$ & Primary   & 91.5 &  1.4 & 36.2 & 95.2 & 3.1  & 1.9  & 49.9 \\
                     & $B$ decay &  6.7 & 46.7 & 29.6 &  3.1 & 75.3 & 10.6 & 22.8 \\
                     & $D$ decay &  1.8 & 51.9 & 34.2 &  1.7 & 21.6 & 87.5 & 27.2 \\
\hline
$500\,\mathrm{GeV}$  & Primary   & 93.7 &  2.6 & 35.3 & 97.4 & 4.9  & 4.0  & 48.5 \\
                     & $B$ decay &  4.6 & 47.3 & 29.8 &  1.8 & 72.3 & 13.5 & 24.5 \\
                     & $D$ decay &  1.7 & 50.1 & 34.9 &  0.8 & 22.9 & 82.5 & 27.0 \\ 
\hline
\end{tabular}
\vspace*{2ex}
\caption{\textit{Percentages of tracks assigned to the reconstructed primary,
secondary and tertiary vertex and of tracks not contained in any vertex 
(labelled ``iso'') which originate from the IP, the $B$ or the $D$ decay
at MC level, for $b$ jets.}}
\label{TableTrackToVertexPuritybJets}
\end{center}
\end{table}
%
%
\begin{table}[htb]
\begin{center}
\begin{tabular}{||cc|ccc|cccc||}
\hline
\multicolumn{2}{||c|}{Monte Carlo} &
\multicolumn{7}{c||}{Reconstructed track-vertex association} \\
\multicolumn{2}{||c|}{track origin} &
\multicolumn{3}{c|}{Two vertex case} &
\multicolumn{4}{c||}{Three vertex case} \\ 
\hline
 & & $\ \ $pri$\ \ $ & sec & iso & $\ \ $pri$\ \ $ & sec & ter & iso \\
\hline
$91.2\,\mathrm{GeV}$ & Primary   & 95.9 &  6.3 & 77.0 & 96.8 & 30.5 & 29.1 & 77.7 \\ 
                     & $D$ decay &  4.1 & 93.7 & 23.0 &  3.2 & 69.5 & 70.9 & 22.3 \\
\hline
$500\,\mathrm{GeV}$  & Primary   & 97.5 &  9.0 & 73.1 & 96.7 & 20.2 & 51   & 73.6 \\
                     & $D$ decay &  2.5 & 91.0 & 26.9 &  3.3 & 79.3 & 48.7 & 26.4 \\ 
\hline
\end{tabular}
\vspace*{2ex}
\caption{\textit{Percentages of tracks assigned to the reconstructed primary,
secondary and tertiary vertex and of tracks not contained in any vertex 
(labelled ``iso'') which originate from the IP or the $D$ decay
at MC level, for $c$ jets. Fractions missing from 100$\,$\% are due to $b$ jets
arising from gluon splitting.}}
\label{TableTrackToVertexPuritycJets}
\end{center}
\end{table}
%
%
\begin{table}[htb]
\begin{center}
\begin{tabular}{||cc|ccc|cccc||}
\hline
\multicolumn{2}{||c|}{Monte Carlo} &
\multicolumn{7}{c||}{Reconstructed track-vertex association} \\
\multicolumn{2}{||c|}{track origin} &
\multicolumn{3}{c|}{Two vertex case} &
\multicolumn{4}{c||}{Three vertex case} \\ 
\hline
 & & $\ \ $pri$\ \ $ & sec & iso & $\ \ $pri$\ \ $ & sec & ter & iso \\
\hline
$91.2\,\mathrm{GeV}$ & $b$ jets  & 50.2 & 35.8 & 13.9 & 39.6 & 28.9 & 23.6 &  8.0 \\
                     & $c$ jets  & 65.1 & 27.3 &  7.6 & 50.6 & 22.0 & 20.5 &  7.0 \\
\hline
$500\,\mathrm{GeV}$  & $b$ jets  & 57.6 & 27.3 & 15.1 & 52.5 & 20.6 & 16.3 & 10.6 \\
                     & $c$ jets  & 74.3 & 17.5 &  8.3 & 65.4 & 13.7 & 13.0 &  7.9 \\
\hline
\end{tabular}
\vspace*{2ex}
\caption{\textit{Percentages of tracks assigned to each type of reconstructed
vertex or left unassigned, for $b$ and $c$ jets.}}
\label{TableTrackToVertexPurityPercentagesInClasses}
\end{center}
\end{table}
In addition to the precision with which the decay lengths can be
reconstructed, important for many physics studies, it was investigated to
which extent the track content of the reconstructed vertices is correct.
For each type of reconstructed vertex --- the primary, the secondary and,
if available, the tertiary vertex --- the average percentage of tracks
that originated from the corresponding MC decay vertex was determined, 
yielding the purity of the track content of each type of vertex.
These purities are given in Table \ref{TableTrackToVertexPuritybJets} for 
$b$ jets%
\footnote{Note that the normalisation is chosen differently from the
earlier SLD table \cite{Jackson:1996sy}: percentages are normalised to 
the total number of tracks in each type of {\it reconstructed vertex}, 
whereas for the former result percentages were given with respect to the 
total number of tracks in each type of {\it MC vertex}.}%
and in Table \ref{TableTrackToVertexPuritycJets} for $c$ jets, 
separately for the cases that exactly two and exactly three vertices 
were reconstructed by ZVRES.
The percentages of tracks that were found to be contained in each vertex
category or left unassigned are given in Table 
\ref{TableTrackToVertexPurityPercentagesInClasses}.
The study was performed at both $\sqrt{s} = 91.2\,\mathrm{GeV}$ and at
the initial maximum energy of the ILC, $\sqrt{s} = 500\,\mathrm{GeV}$.
The best assignment of tracks to vertices, corresponding to the highest
purities, was obtained for $c$ jets with two reconstructed vertices and
for $b$ jets with three reconstructed vertices.
This is understandable given the fact that if both the $b$ and the 
subsequent charm decay result in multi-prong vertices, this corresponds
to a cleaner topology that can be more easily reconstructed.
If a smaller number of vertices is found in $b$ jets, this can indicate
either that one of the decay vertices is one-pronged and thus cannot
be found by ZVRES, or that the decay length of one of the heavy flavour 
hadrons is so short that its decay vertex cannot be resolved from the 
preceding vertex in the decay chain, and that, for example, the
secondary vertex found by ZVRES contains some tracks that actually 
originated from the $D$ decay and some from the $B$ decay.
Both effects can result in a misassignment of tracks by ZVRES and hence
to a reduced purity for vertices in this category of jets.
Similarly, the sample of $c$ jets with three reconstructed vertices will
contain a higher rate of fake vertices than the $c$ jet sample with two
reconstructed vertices, again corresponding to a higher confusion in the
track-to-vertex association.

Comparing the results at $91.2$ and at $500\,\mathrm {GeV}$ centre-of-mass
energy, an increase in the available energy results in an increase in the 
number of tracks originating from the primary vertex.
It is thus expected that the percentage of tracks assigned to the primary
vertex by ZVRES increases, as seen in 
Table \ref{TableTrackToVertexPurityPercentagesInClasses}.
This increased multiplicity of IP tracks, in combination with jets becoming
more collimated at higher energies, makes vertex finding more challenging,
even though these effects are partly compensated by increasing decay lengths.
The net effect is an increased confusion in the track-to-vertex assignment,
which is most pronounced for the tertiary vertex in three-vertex $c$ jets,
for which the percentage of IP tracks increases from about $30\,\%$ at
$\sqrt{s} = 91.2\,\mathrm{GeV}$ to about $50\,\%$ at 
$\sqrt{s} = 500\,\mathrm{GeV}$.
The relatively large changes of these numbers with centre-of-mass energy
indicate that these effects will need to be studied in more detail in the
future; in particular it would be worth investigating if performance at 
higher energy can be improved by adjusting the energy dependence of the 
ZVRES parameters.

{The performance studies of the vertex reconstruction at a
centre-of-mass energy of $91.2\,\mathrm{GeV}$ allow a direct
comparison with results obtained at the SLD experiment, with vertex
detectors VXD2 \cite{Jackson:1996sy} and VXD3 \cite{SLD:1997}.  The
improved angular coverage, point resolution and reduced material
budget envisaged for the ILC vertex detector are expected to result in
significant improvements in performance over SLD. Indeed, the vertex
finding efficiency for the ILC vertex detector model, shown in
Figure~\ref{FigureZVTOPProbAndEff}, is clearly improved compared to
the earlier SLD results, increasing rapidly at low decay lengths and
reaching an average value of $89\,\%$ in the plateau region above
decay lengths of $1\,\mathrm{mm}$.  In comparison, for SLD-VXD3, the
plateau was only reached for decay lengths of about $2\,\mathrm{mm}$,
with the efficiency above that value being about $80\,\%$
\cite{SLD:2000}.}

\section{A flavour tag procedure based on neural networks}
\label{SectionFlavourTag}

{The LCFIVertex software package contains a neural network based jet
flavour tag modeled closely after an earlier Fortran-based
implementation by R.~Hawkings~\cite{Hawkings:2000}. A distinctive
feature of this approach is a separate treatment of jets with and
without non-IP vertices. Beyond being a full re-implementation of this
algorithm in C++, the LCFIVertex package features a high degree of
flexibility concerning neural network architecture, choice of input
variables and a tool to determine the relevance of individual neural
network inputs. The actual neural network setup and parameters for
LCFIVertex are defined by external files loaded at run time. These
files can be maintained and distributed independently of the
LCFIVertex code, allowing the provision of central repositories of
neural network files tuned to specific detector models and/or
centre-of-mass energies.}

\subsection{Determination of true jet flavour, hadron and quark charge
from MC}
\label{SubsectionMCTruth}

In order to define performance measures for flavour tag and vertex/quark
charge, the true jet flavour and the charge of the leading hadron and of
the heaviest quark contained in it need to be known for comparison.
A dedicated part of the LCFIVertex package implements the following
algorithm to extract this information from the event record of the MC
generator that is included in the LCIO event:
the event is searched for all hadrons containing $b$ or $c$ quarks.
These hadrons are assigned to the reconstructed jet closest in angle, 
with the possibility of assigning more than one heavy hadron per jet.
From the hadrons assigned to a given jet, the one appearing earliest
in the MC decay chain is selected and the jet assigned the flavour of
the heaviest quark contained in it as true jet flavour.

For two-jet events, the true jet flavour is clearly defined and 
easily obtained from this procedure.
In multi-jet events, where the angular distance between jets is 
smaller and the assignment of tracks to jets sometimes becomes
ambiguous, there may be cases of tracks corresponding to hadrons from
the same parton shower being assigned to different jets, with the 
concept of a ``jet flavour'' becoming less clearly defined.
However, these are exceptions related to the general difficulties
of jet finding in such events, and the algorithm described above
yields good results also for most jets in multi-jet events.

\subsection{Observables sensitive to jet flavour}
\label{SubsectionFlavourTagInputs}

Many of the variables most sensitive to jet flavour are only 
defined for jets in which non-IP vertices have been found.
{Therefore, different sets of observables are used for jets containing
one and jets containing more than one vertex.}

%
\begin{figure}[hp]
\begin{center}
\begin{tabular}{cc}
\includegraphics[width=0.50\columnwidth]{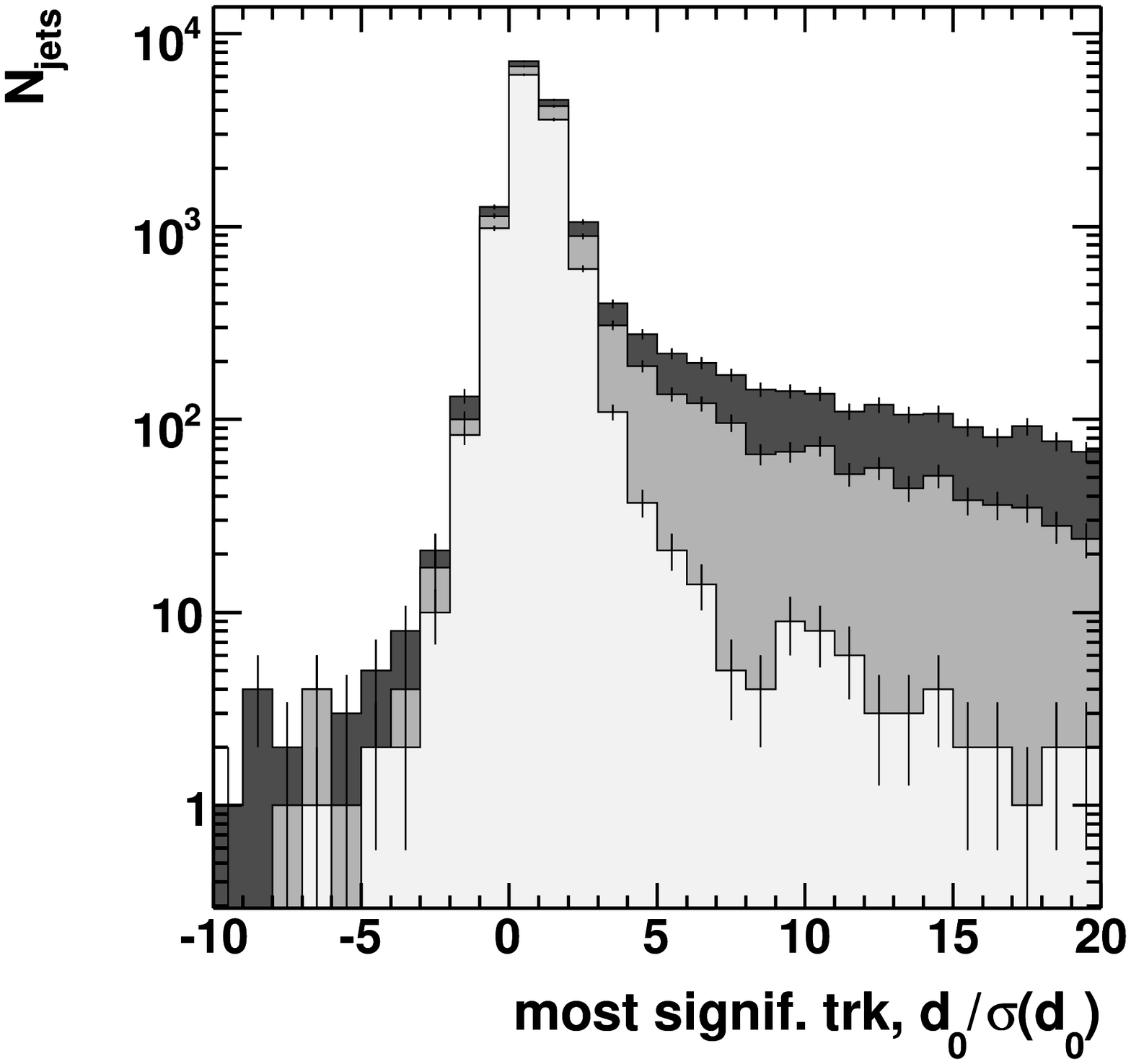} &
\includegraphics[width=0.50\columnwidth]{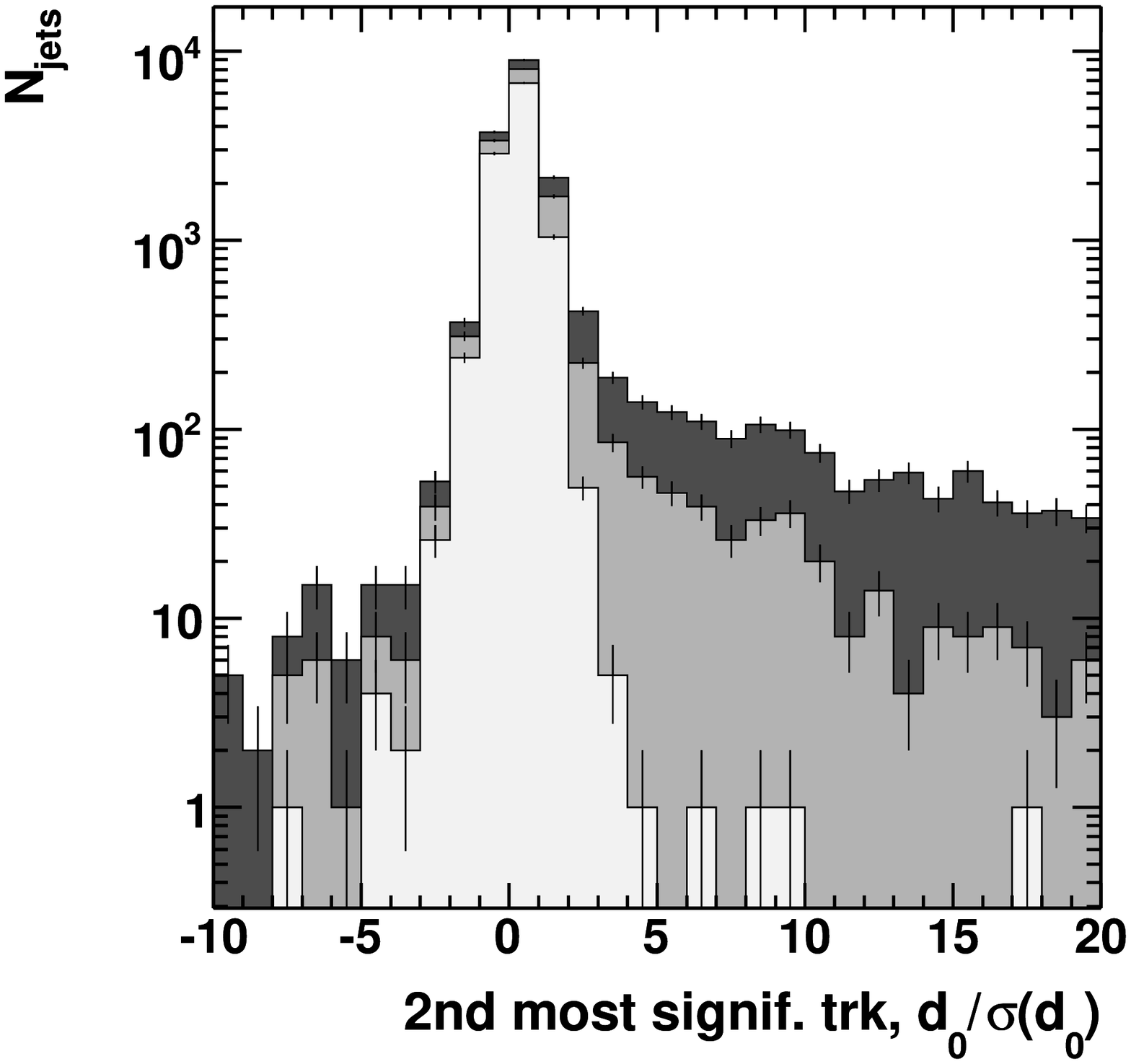}\\[-35ex]
\hspace*{12em}(a) & \hspace*{12em}(b) \\[30ex]
\includegraphics[width=0.50\columnwidth]{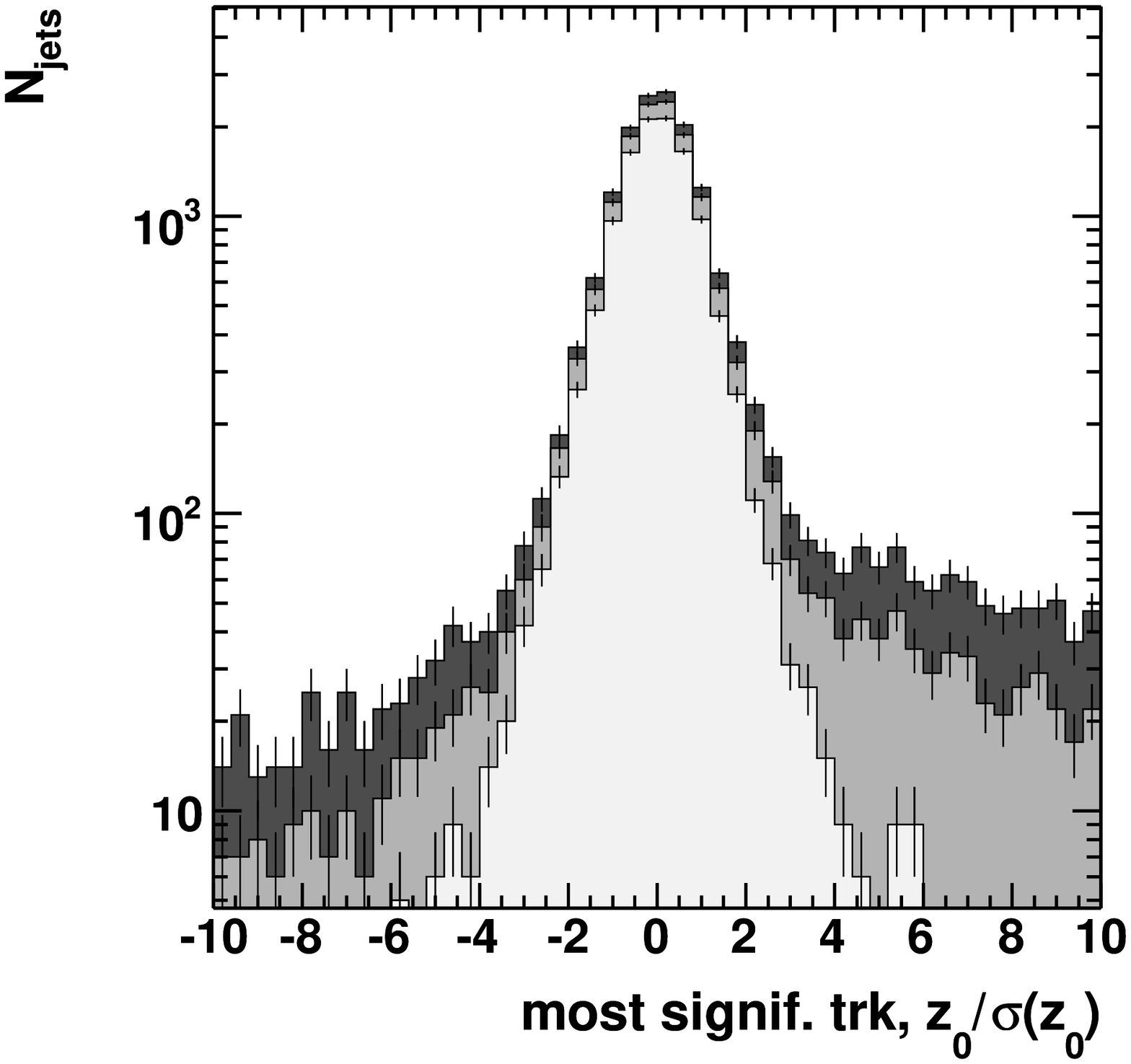} &
\includegraphics[width=0.50\columnwidth]{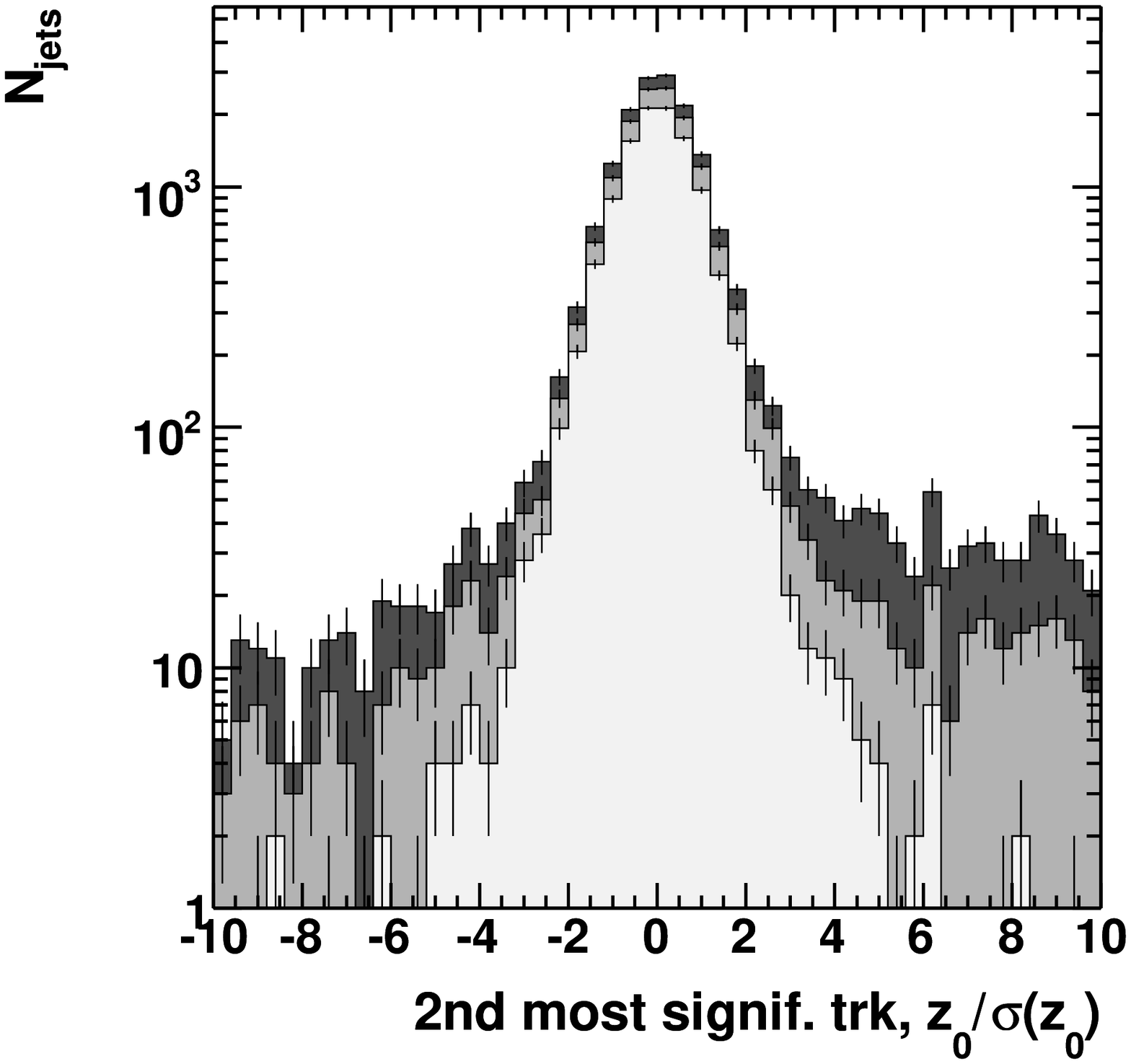}\\[-35ex]
\hspace*{12em}(c) & \hspace*{12em}(d) \\[30ex]
\includegraphics[width=0.50\columnwidth]{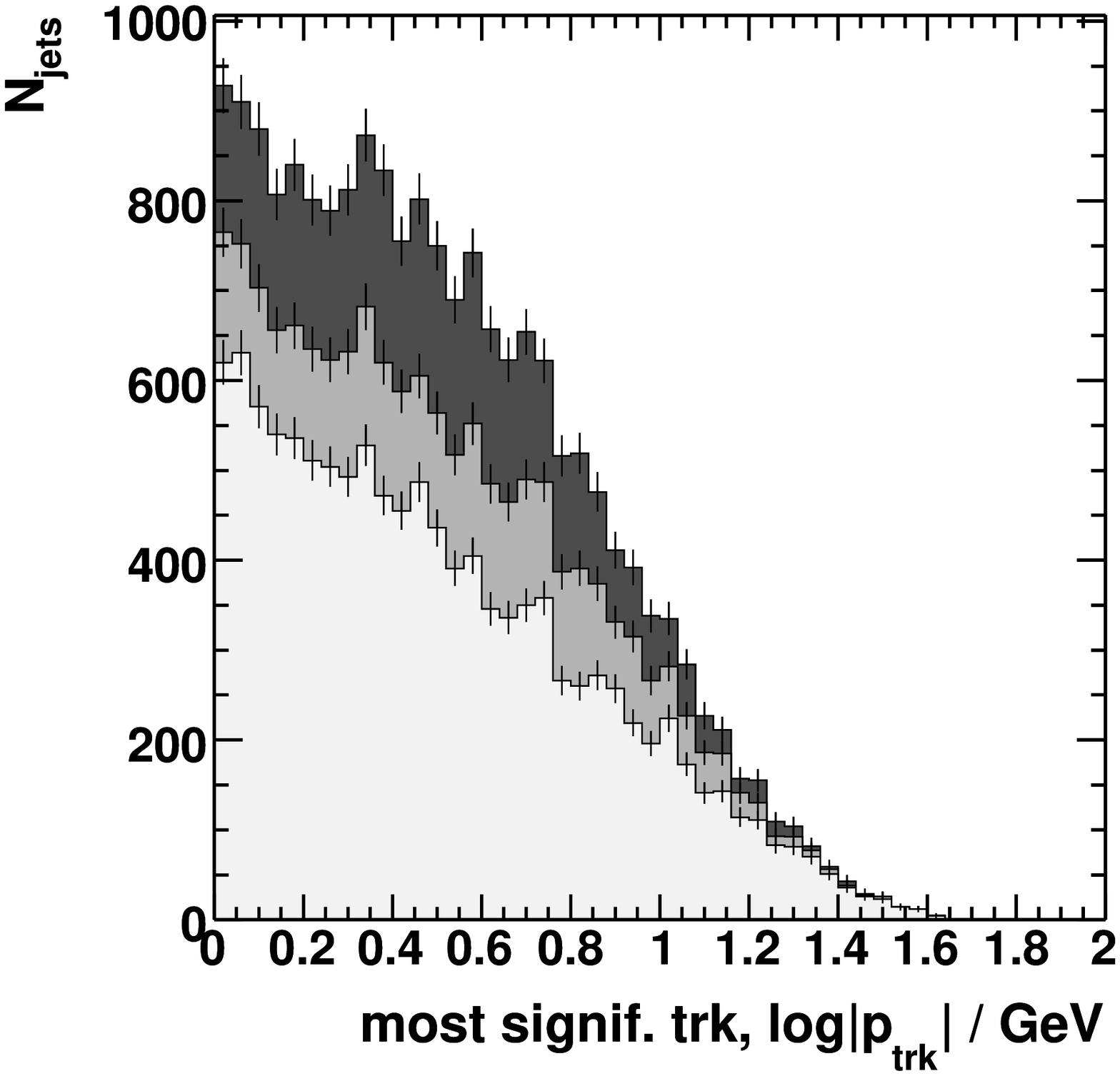} &
\includegraphics[width=0.50\columnwidth]{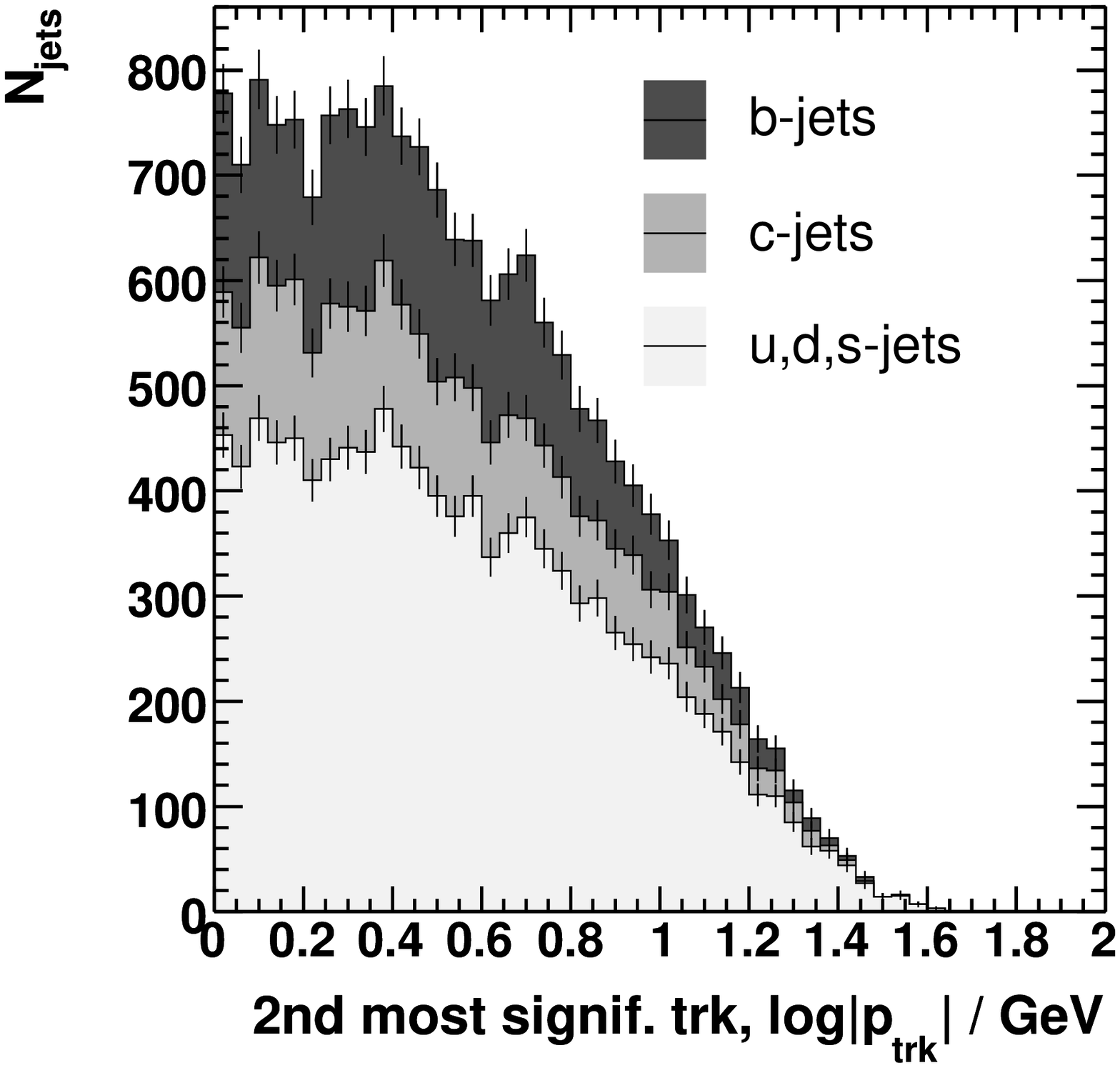}\\[-35ex]
\hspace*{12em}(e) & \hspace*{12em}(f) \\[30ex]
\end{tabular}
\caption{\textit{Flavour tag inputs based on the most [plots (a), (c), (e)]
and second-most [plots (b), (d), (f)] significant track in the jet.
Shown are the impact parameter significance in $R$-$\phi$ in (a), (b),
the impact parameter significance in $R$-$z$ in (c), (d) and the track
momentum in (e), (f).}}
\label{FigureFlavourTagInputsOneVertex}
\end{center}
\end{figure}
In the case that only one vertex --- the event vertex --- has been
found, the input jet is searched for the two tracks of highest 
impact parameter significance%
\footnote{The impact parameter of a track is defined as the distance
between the track's point of closest approach to the IP and the IP.
The impact parameter significance is the impact parameter divided by
its uncertainty.}
in the $R$-$\phi$ plane.
These are referred to as the most significant and the second-most
significant track in what follows.
For finding these two tracks, separate minimum momentum cuts
$p_{\mathrm{trk,NL,min}}$ and $p_{\mathrm{trk,NL-1,min}}$ are applied
for tracks with hits on all $N_{L}$ vertex detector layers or with
hits on only $N_{L}-1$ layers, respectively.
The momenta $|p_{\mathrm{trk}}|$ and impact parameter significances 
of these tracks in the $R$-$\phi$ and $R$-$z$ planes are used as input 
for the flavour tag.
A single track of high impact parameter significance may indicate that 
the jet under consideration is a charm jet with the leading $D^{\pm}$
having decayed to a single charged track (``one-prong'' decay), which
is expected for $\approx 40\,\%$ of all $D^{\pm}$ decays \cite{PDG:2008}.
The observables obtained from the second-most significant track help
distinguish between $c$ and $b$ jets, for which it is more likely that
two tracks of high impact parameter significance are found, typically with
one resulting from the decay of the leading hadron and one from the decay of 
the charmed hadron produced in that decay.
Fig.~\ref{FigureFlavourTagInputsOneVertex} shows distributions of the 
inputs for the most significant and second-most significant tracks, 
separately for $b$, $c$ and light flavour jets. 
{It is worth noting that a small positive tail of impact parameter
significances is observed in uds jets, where the largest positive
impact parameter significance is occasionally contributed by a single
misreconstructed track in the jet. These tracks lead to an important
background for the identification of one-prong charm decays.}

Further information is contained in the ``joint probability'' for all tracks
to originate from the primary vertex, as introduced by ALEPH 
\cite{ALEPH:JointProb:1993}.
The first implementation of this variable for an ILC detector is described
elsewhere \cite{Hawkings:2000}.
Two joint probability variables are calculated from the impact parameter
significances in $R$-$\phi$ and in $R$-$z$ of all the tracks in the jet that
pass the specific selection criteria as detailed below.
The distribution $f(x)$ of unsigned impact parameter significances for IP tracks is
assumed to be known; it can be determined from the data, as described in
Appendix \ref{AppendixJointProbability}.
The probability of an IP track having an impact parameter significance of
$b/\sigma_{b}$ or larger is given by
\[
P_{i} = \frac{\int_{b/\sigma_{b}}^{\infty} f(x) dx}{\int_{0}^{\infty} f(x) dx}\ \ \ .
\]
For a set of $N$ tracks the probability that all $N$ tracks originate from 
the IP is 
\[ 
P_{J} = y \sum_{k=0}^{N-1} \frac{\left( -\ln y\right)^{k}}{k!}, \ \ \
y = \prod_{i} P_{i}\ \ \ .
\]
The joint probability is the observable $P_{J}$, calculated for the set of
tracks that pass the track selection cuts described in Section 
\ref{SubsectionTrackSelection} as well as an upper cut on impact parameter of 
$5\,\mathrm{mm}$ and on impact parameter significance of $200$.
It is calculated separately for the $R$-$\phi$ and the $R$-$z$ impact parameter
significances.

\begin{figure}[htb]
\begin{center}
\begin{tabular}{cc}
\includegraphics[width=0.50\columnwidth]{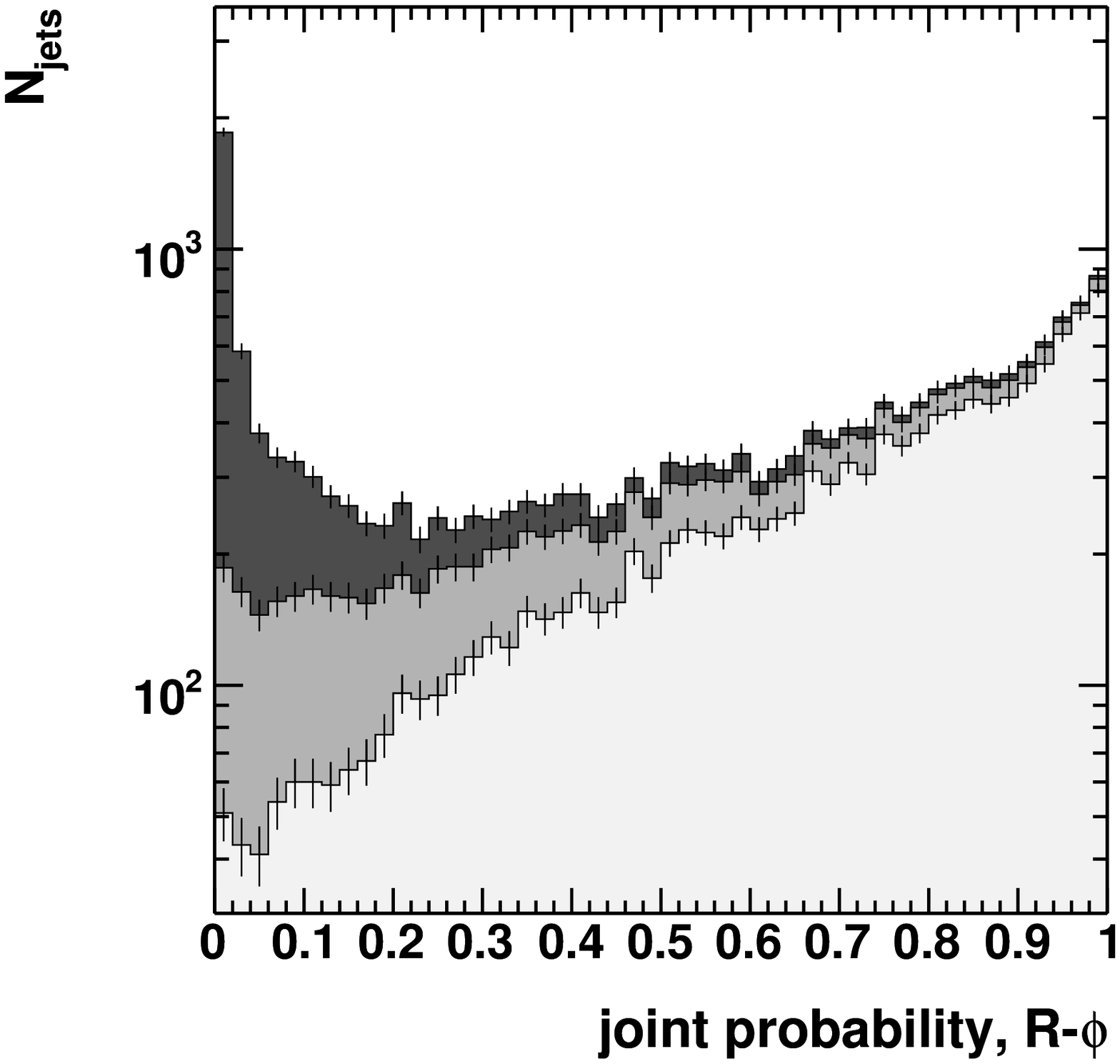} &
\includegraphics[width=0.50\columnwidth]{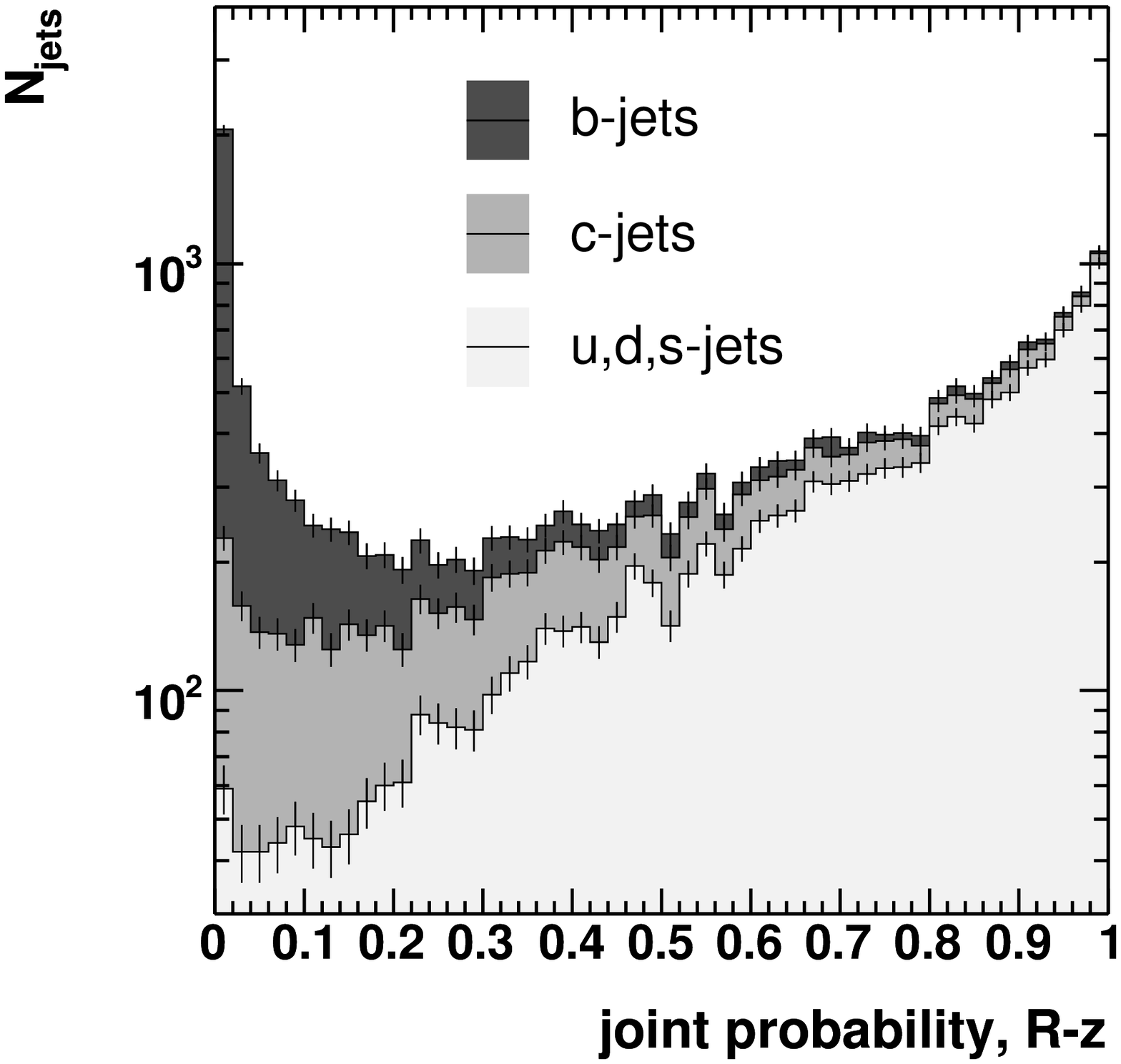}\\[-35ex]
\hspace*{12em}(a) & \hspace*{12em}(b) \\[30ex]
\end{tabular}
\caption{\textit{Joint probability for all tracks in the jet passing the
track selection cuts to originate from the primary vertex.
This probability is calculated separately from (a) the $R$-$\phi$ impact
parameter significances and (b) the impact parameter significances in
$R$-$z$.}}
\label{FigureJointProbability}
\end{center}
\end{figure}
As can be seen from the resulting $P_{J}$ distributions shown in 
Fig.~\ref{FigureJointProbability}, light quark jets tend to have values 
closer to $1$, while the distributions for $b$ and $c$ jets peak at zero.

In the case that more than one vertex is found, observables derived from these
additional vertices provide a more powerful means to distinguish between $b$,
$c$ and light quark jets.
The following set of eight variables is used in that case:
\begin{itemize}
\item{The decay length and decay length significance of the vertex with the
largest decay length significance in three dimensions with respect to the IP;}
\item{The momentum $|p|$ of the set of tracks assigned to the decay chain
(see below);}
\item{The $p_{T}$-corrected vertex mass, calculated as described below;}
\item{The number $N_{\mathrm{trk,vtx}}$ of tracks in all non-primary vertices;}
\item{The secondary vertex probability of the tracks assigned to the
decay chain; a new vertex fit is performed using these tracks and the
probability calculated from the fit $\chi^{2}$;}
\item{The joint probability in $R$-$\phi$ and in $R$-$z$ as described above.}
\end{itemize}
\begin{figure}[htb]
\begin{center}
\includegraphics[width=0.5\columnwidth,clip=true]{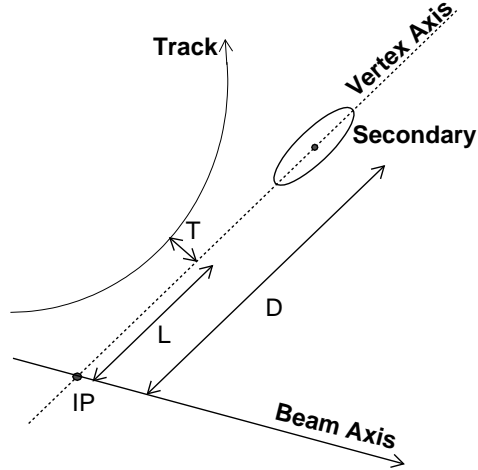}
\caption{\textit{Schematic showing the definition of distances $L$, $D$
and $T$ used in the selection of tracks for vertex mass determination.}}
\label{FigureLByD}
\end{center}
\end{figure}
In addition to the general track selection for the calculation of the
flavour tag variables, a special track selection is applied for the 
variables $M_{\mathrm{Pt}}$, $|p|$ and the secondary vertex probability
as follows:
the ``seed vertex'', i.e.~the vertex furthest away from the IP, is used
to define the distances $D$ between this vertex and the IP, and $L$ as
shown in Fig.~\ref{FigureLByD}.
The ``vertex axis'' is the straight line connecting the seed vertex 
with the IP.
For each track, the point of closest approach to the vertex axis is 
projected onto the vertex axis and $L$ defined as the distance of the 
resulting point on the vertex axis from the IP.
Tracks with $0.18 < L/D < 2.5$ and a transverse distance $T$ of the
point of closest approach from the vertex axis below $1.0\,\mathrm{mm}$
are attached to the decay chain and used in the calculation.
{Note that in the default configuration all tracks from all vertices except the primary vertex are included automatically. Optionally, the track attachment cuts can be applied also
to the tracks in the seed vertex.}

The momentum $|p|$ is the modulus of the vector sum of all decay
chain track momenta.
The secondary vertex probability is found by fitting a common vertex
to these tracks and calculating the probability from the $\chi^{2}$
value of this fit in the same way as for the ZVKIN vertex finder, 
see Section \ref{SubsectionZVKIN}.
For the secondary vertex probability the number of tracks in the
decay chain is required to exceed the value $N_{\mathrm{trks,min}}$
and the normalised fit-$\chi^{2}$ is required to be below a user-settable
value: $\chi^{2}/\sqrt{\mathrm{ndf}} < \chi^{2}_{\mathrm{norm,max}}$,
where $\mathrm{ndf}$ is the number of degrees of freedom. For jets
that do not meet these requirements, the probability is set to 0,
to lower the risk of such jets leaking into the heavy flavour samples.

For the calculation of the $P_{T}$-corrected vertex mass 
$M_{\mathrm{Pt}}$, first the vertex mass before correction
$M_{\mathrm{Vtx}}$, is obtained from the decay chain tracks,
assigning pion mass to each track.
With $\theta_{\mathrm{Vtx}}$ the angle between the seed vertex axis
as given by the vertex position and the vertex momentum with 
respect to the IP, it is required that
$p^{2}\cdot(1 - \cos^{2}\theta_{\mathrm{Vtx}}) \leq 
w_{\mathrm{Pt,max}}\cdot M^{2}_{\mathrm{Vtx}}$, 
where the factor $w_{\mathrm{Pt,max}}$ is a user-defined LCFIVertex parameter of 
default value 3.
This cut ensures that cases in which both $\theta_{\mathrm{Vtx}}$
and $|p|$ are large are excluded from the correction procedure
to reduce the risk of fake vertices being assigned a large
correction and subsequently affecting the flavour tag.
Jets failing this cut are assigned an $M_{\mathrm{Pt}}$ value
of $0$.
A conservative estimate of the transverse momentum 
$p_{T}^{\mathrm{Vtx}}$ corresponding to $\theta_{\mathrm{Vtx}}$
and taking the error matrices of the seed axis and the IP into
account, is obtained by iteratively minimising the correction 
term $p_{T}^{\mathrm{Vtx}}$ and recalculating the seed axis
direction.
For this minimisation, the parameter $N_{\sigma,\mathrm{max}}$
determines the permitted extent of the seed axis correction
in units of its uncertainty, larger values of 
$N_{\sigma,\mathrm{max}}$ permitting larger changes.
The $P_{T}$-corrected vertex mass $M_{\mathrm{Pt}}$ is then
defined as 
$M_{\mathrm{Pt}} = \sqrt{M^{2}_{\mathrm{Vtx}} + \left|p_{T}^{\mathrm{Vtx}}\right|^{2}}
                 + |p_{T}^{\mathrm{Vtx}}|$.
Finally, it is required that the correction does not exceed the
uncorrected value by a large factor, 
$M_{\mathrm{Pt}} \leq w_{\mathrm{corr,max}}\cdot M_{\mathrm{Vtx}}$,
where $ w_{\mathrm{corr,max}}$ is a code parameter.

\begin{figure}[hp]
\begin{center}
\begin{tabular}{cc}
\includegraphics[width=0.50\columnwidth]{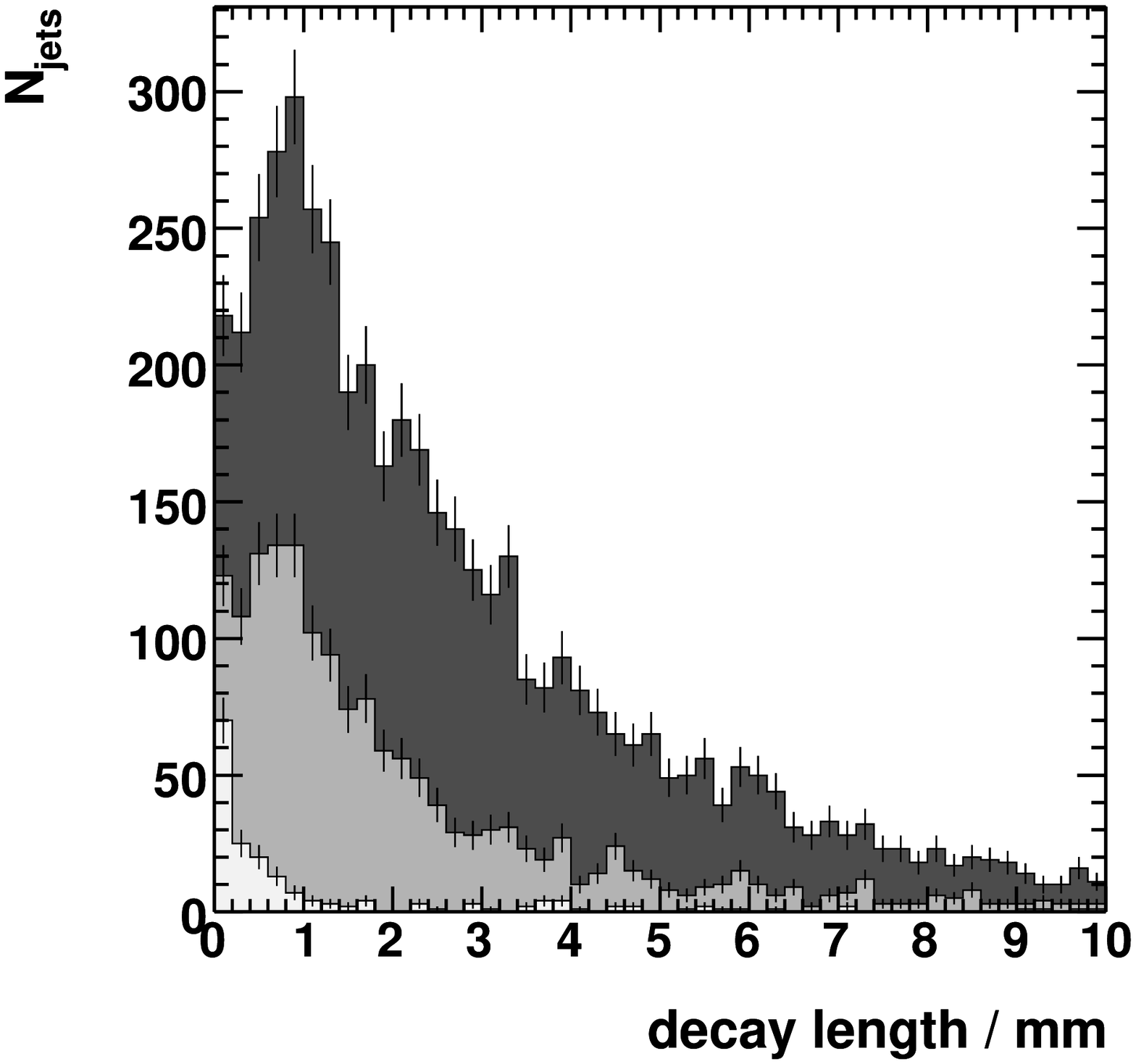} &
\includegraphics[width=0.50\columnwidth]{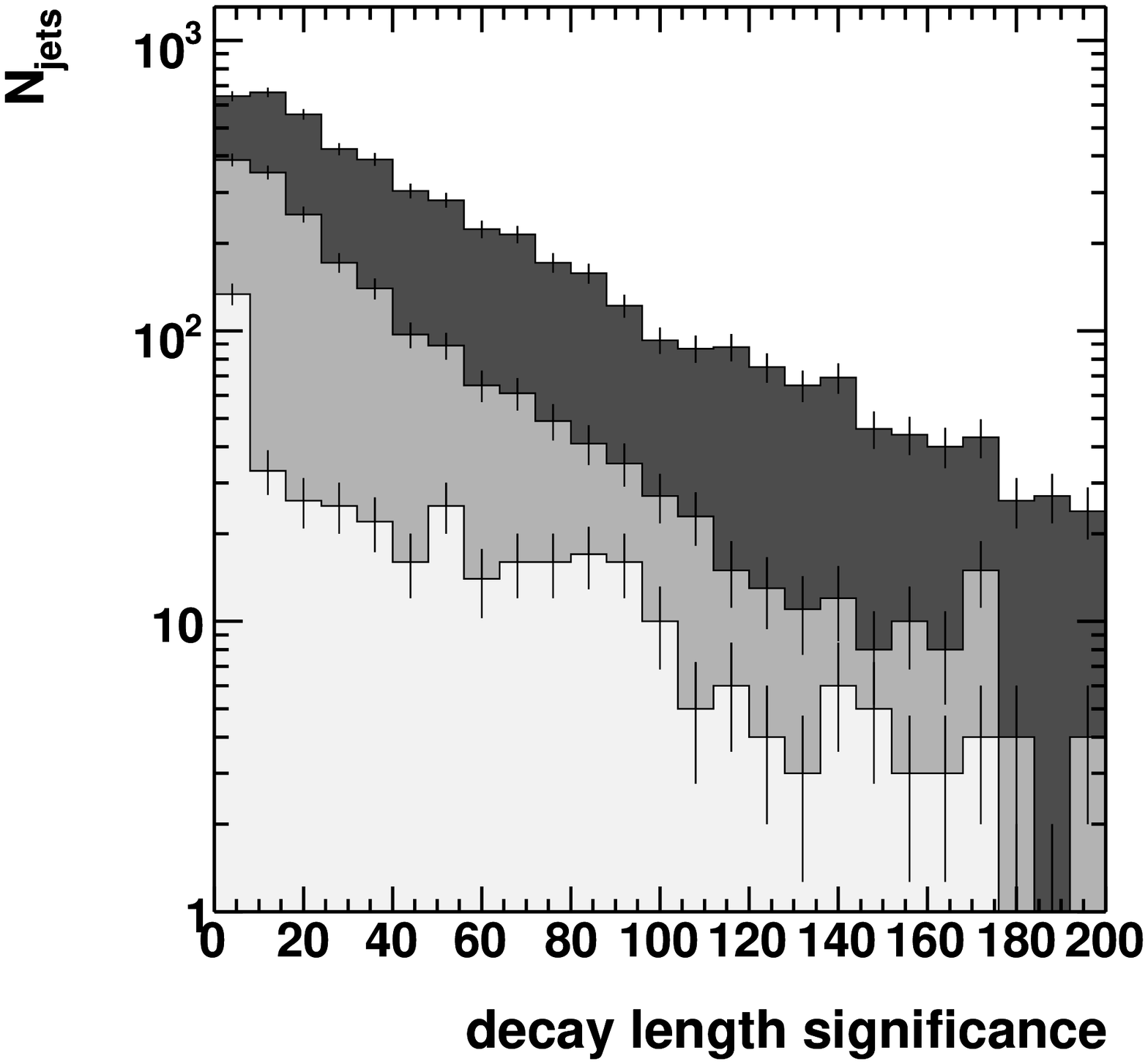}\\[-35ex]
\hspace*{12em}(a) & \hspace*{12em}(b) \\[30ex]
\includegraphics[width=0.50\columnwidth]{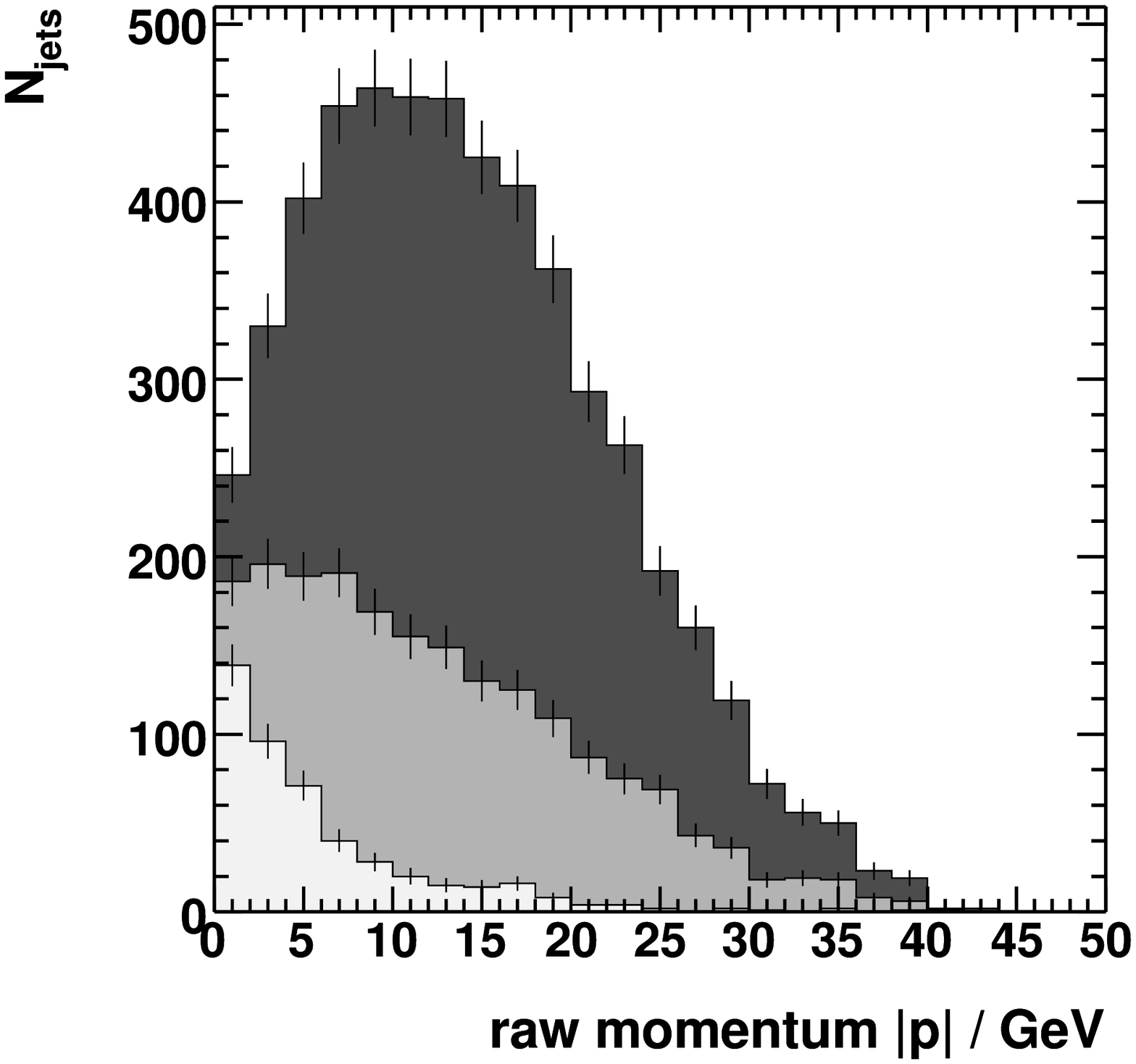} &
\includegraphics[width=0.50\columnwidth]{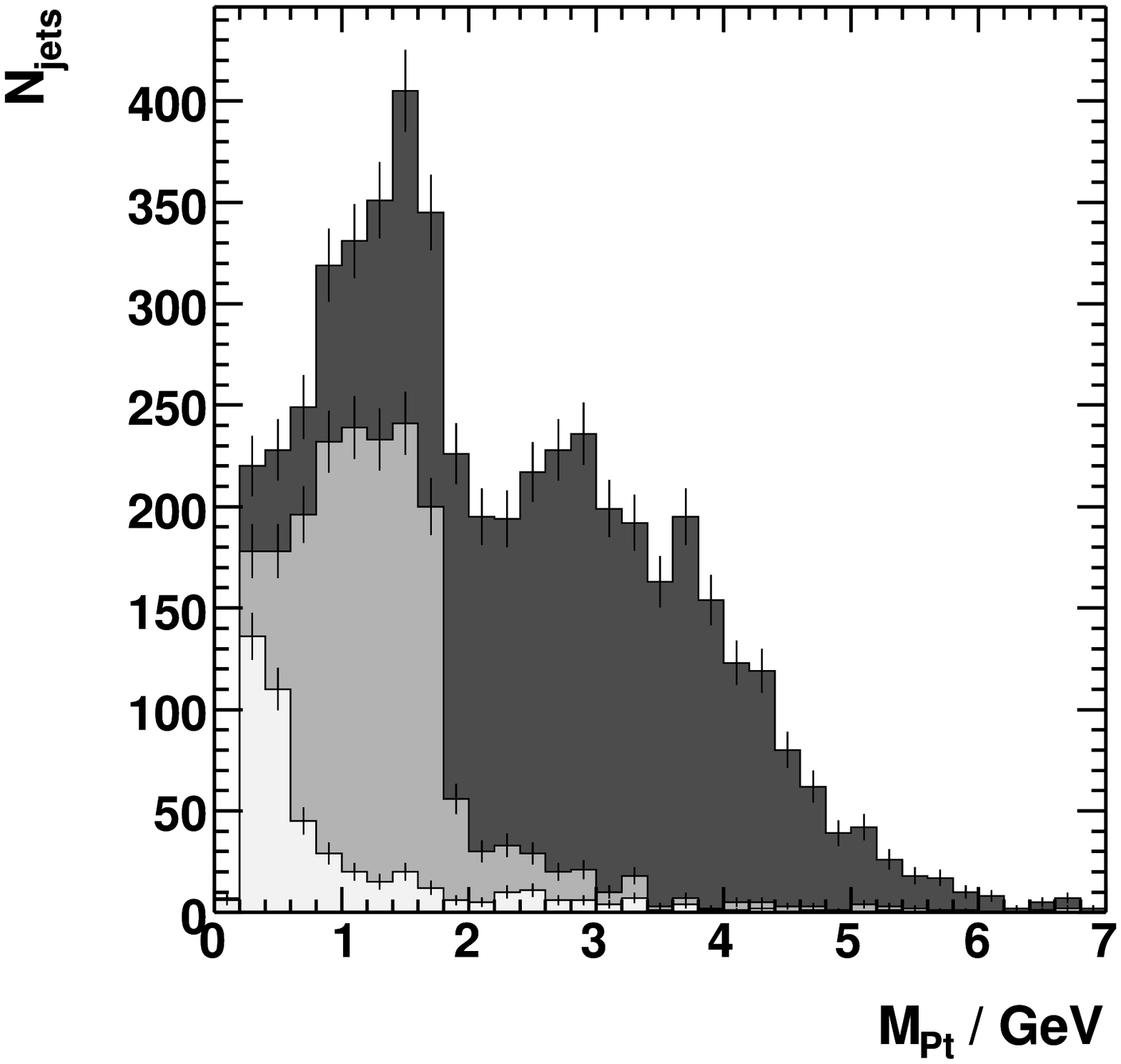}\\[-35ex]
\hspace*{12em}(c) & \hspace*{12em}(d) \\[30ex]
\includegraphics[width=0.50\columnwidth]{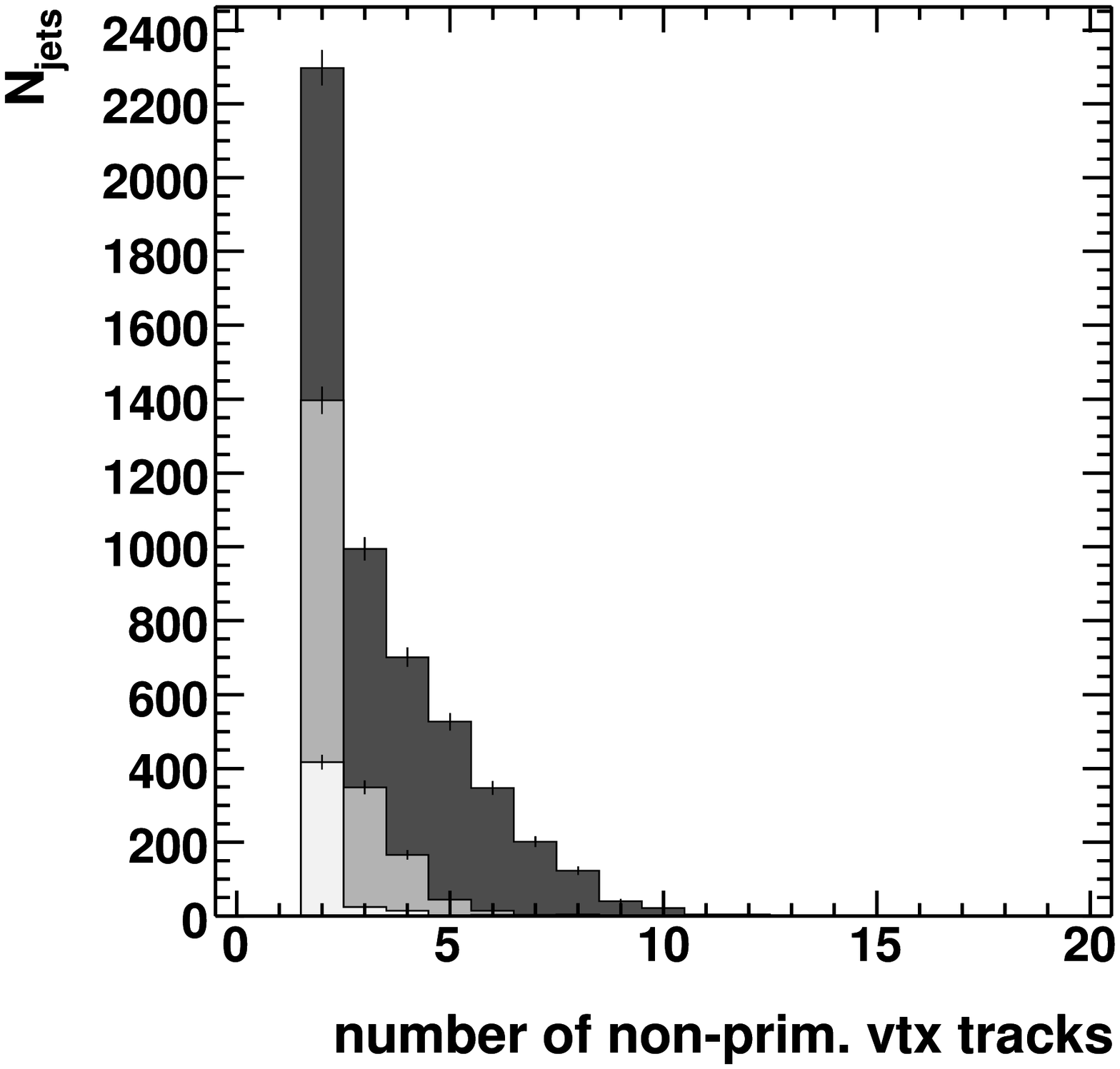} &
\includegraphics[width=0.50\columnwidth]{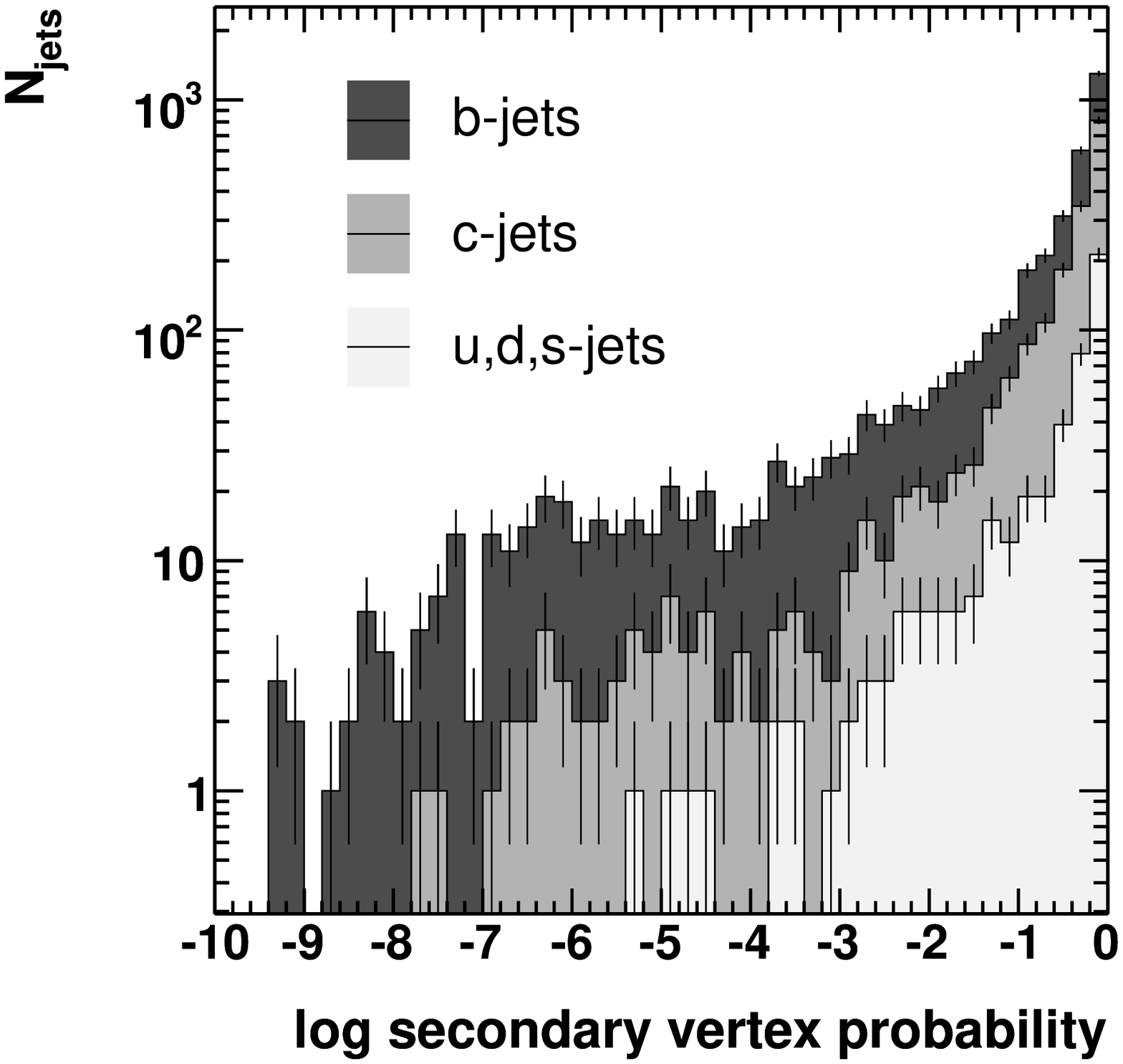}\\[-35ex]
\hspace*{12em}(e) & \hspace*{12em}(f) \\[30ex]
\end{tabular}
\caption{\textit{Flavour tag inputs that are used if at least two vertices
(at least one secondary) are found in the jet.}}
\label{FigureFlavourTagInputsGeqTwoVertices}
\end{center}
\end{figure}
Figure~\ref{FigureFlavourTagInputsGeqTwoVertices} shows the flavour 
tag input variables used for jets for which more than one vertex 
was found, for $b$, $c$ and light quark jets separately.
Some of these variables, such as $M_{\mathrm{Pt}}$, already provide
very good separation of the different jet flavours on their own,
with correlations between the observables, exploited by the neural
network approach, further improving the tagging performance.

Some of the input variables of the flavour tag depend on the 
energy of the input jet.
In order to be able to use the neural networks that are trained 
with jets from the 
$e^{+}e^{-} \rightarrow Z/\gamma \rightarrow q\bar{q}$ events
at the $Z$ resonance for arbitrary energy, the momenta of the 
most and second-most significant track, the decay length
significance and the seed vertex momentum are normalised to the 
jet energy before being fed into the neural nets.

\subsection{Combining flavour-sensitive variables using 
neural networks}
\label{SubsectionNeuralNets}

For heavy flavour tagging, that is the identification of bottom and
charm jets, neural networks are trained such that the target output
provided in the training phase is $1$ for signal jets and $0$ for
background.
The output value of a trained network will be the closer to $1$ 
the more signal-like the values of the input observables.
{The LCFIVertex code is very flexible,} permitting the use
of different input variables, network architecture, node type,
transfer function and training algorithm.

\begin{figure}[htb]
\begin{center}
\begin{tabular}{cc}
\includegraphics[width=0.50\columnwidth]{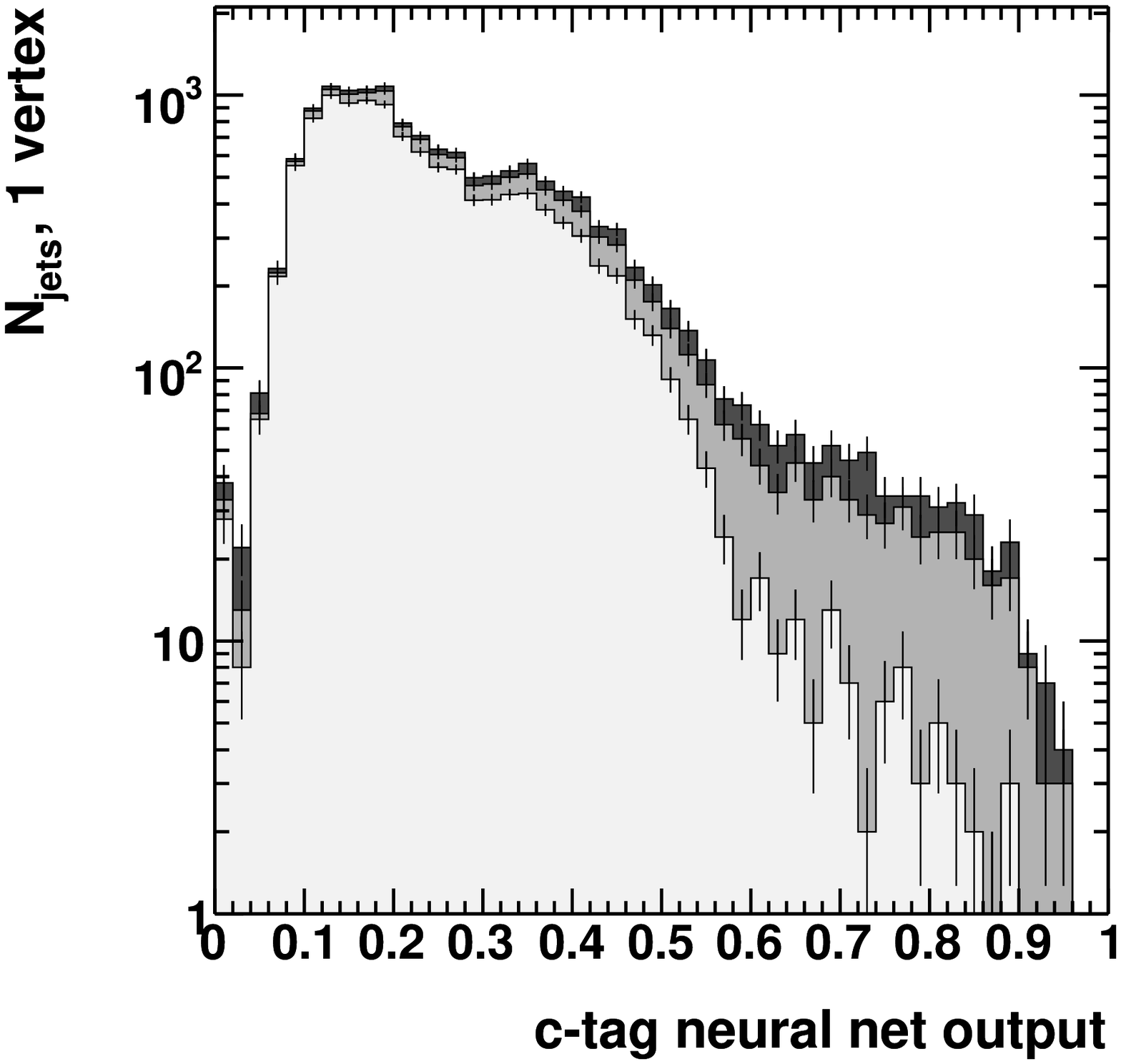} &
\includegraphics[width=0.50\columnwidth]{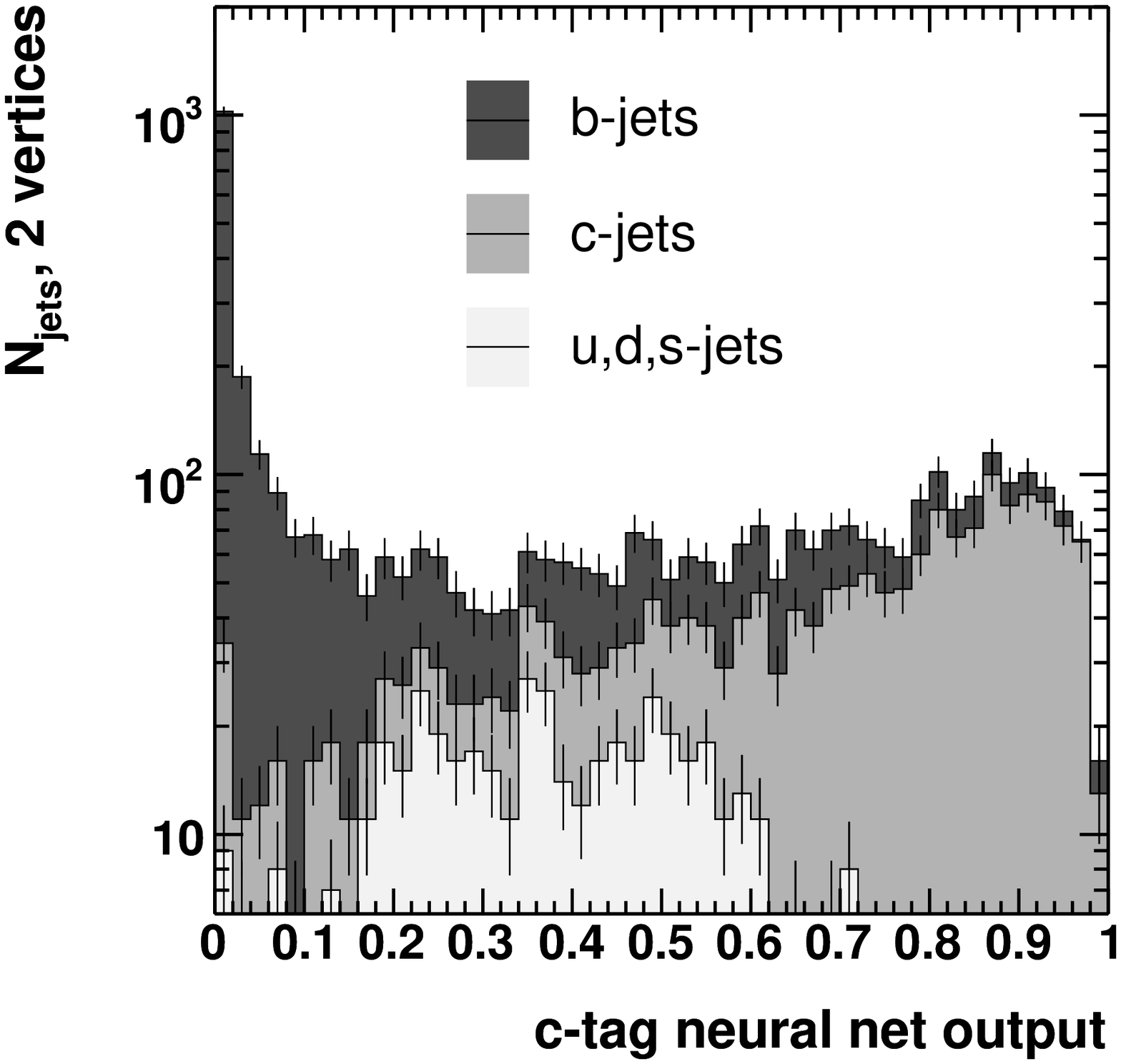}\\[-35ex]
\hspace*{12em}(a) & \hspace*{12em}(b) \\[30ex]
\includegraphics[width=0.50\columnwidth]{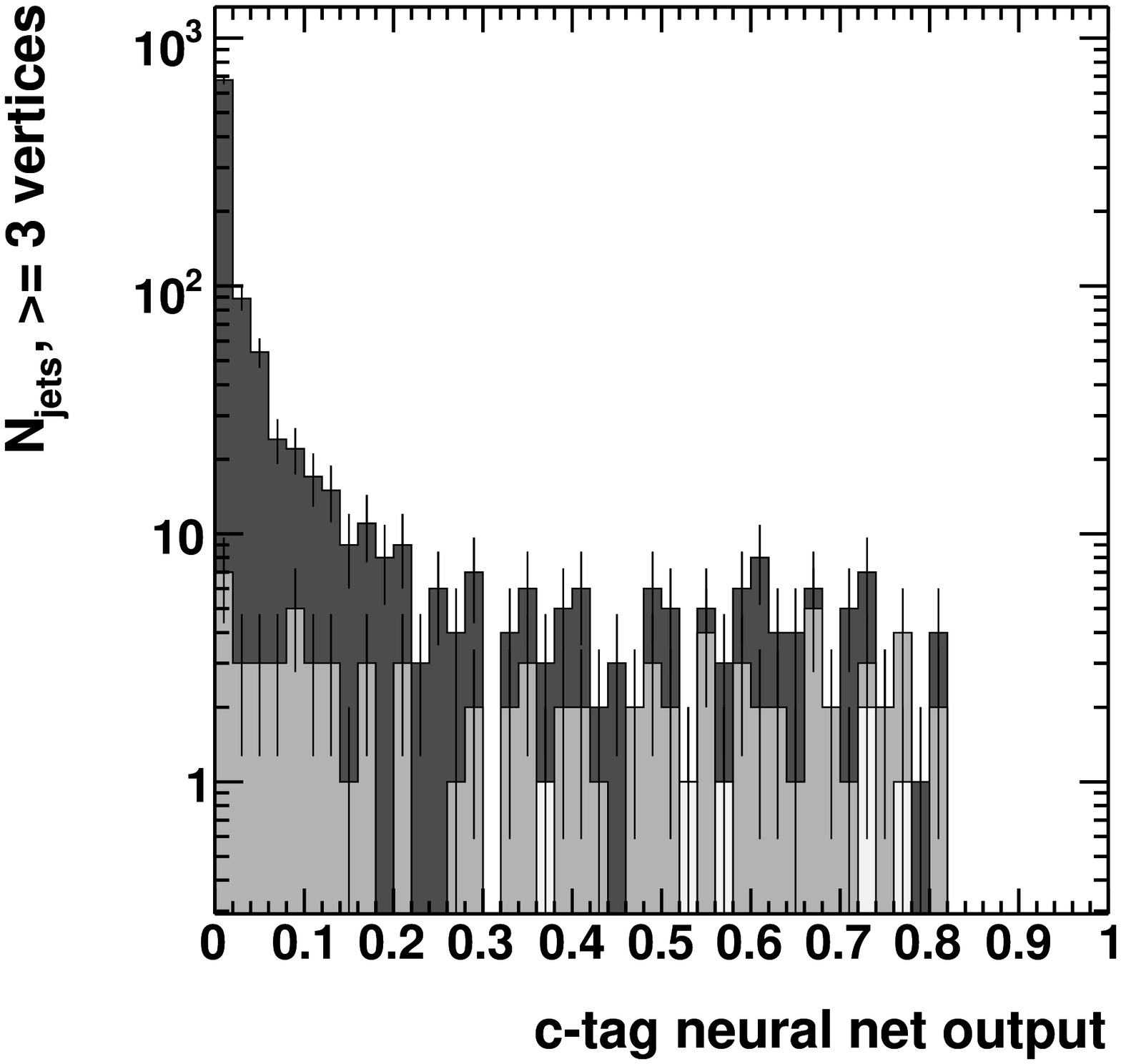} &
\includegraphics[width=0.50\columnwidth]{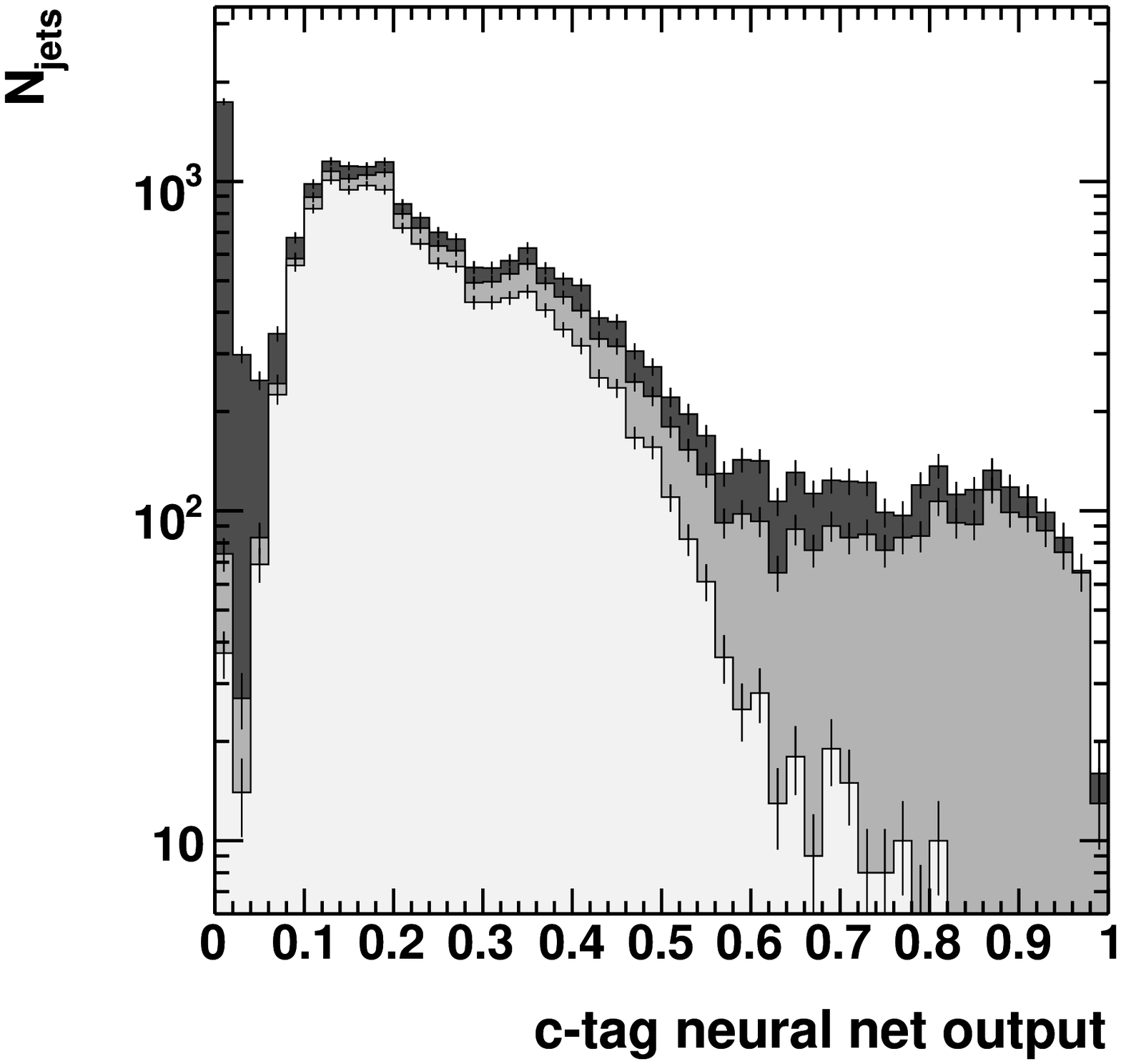}\\[-35ex]
\hspace*{12em}(c) & \hspace*{12em}(d) \\[30ex]
\end{tabular}
\caption{\textit{Output of the neural networks used for charm
tagging. The plots show the outputs for the three separate
networks used in case (a) one, (b) two and (c) three or more vertices
are found in the input jet. In (d), the resulting distribution for
arbitrary number of vertices is shown. }}
\label{FigureCTagOutput}
\end{center}
\end{figure}
%
{By default, each of the networks is a multilayer perceptron with} 8 input nodes, one hidden layer
of 14 tan-sigmoid nodes and one output node and is trained using 
the conjugate gradient back propagation algorithm.
As explained in Section \ref{SubsectionNeuralNets}, the flavour tag 
is based on different observables for jets with one and for jets with 
two or more found vertices. 
Furthermore, for a given jet flavour, the distributions of sensitive
variables are significantly different for jets with two and jets with
three or more vertices, so the ability to distinguish between $b$
and $c$ jets is enhanced by treating these two cases separately.
For each of these categories of one, two or at least three vertices,
three networks are trained, so a complete set consists of nine 
networks altogether: for ``$b$ nets'', the signal provided in the
training phase consists of $b$ jets while $c$ and light flavour jets
form the background.
The ``$c$ nets'' are trained with $c$ jets as signal and $b$ and
light flavour jets as background. 
As for some physics processes the background for the identification of
$c$ jets is known to consist of $b$ jets only, and charm jets are
easier to distinguish from these than from light flavour jets, 
dedicated networks are provided for this case, which are only 
presented $b$ jets as background in the training run.

\begin{figure}[htb]
\begin{center}
\begin{tabular}{cc}
\includegraphics[width=0.50\columnwidth]{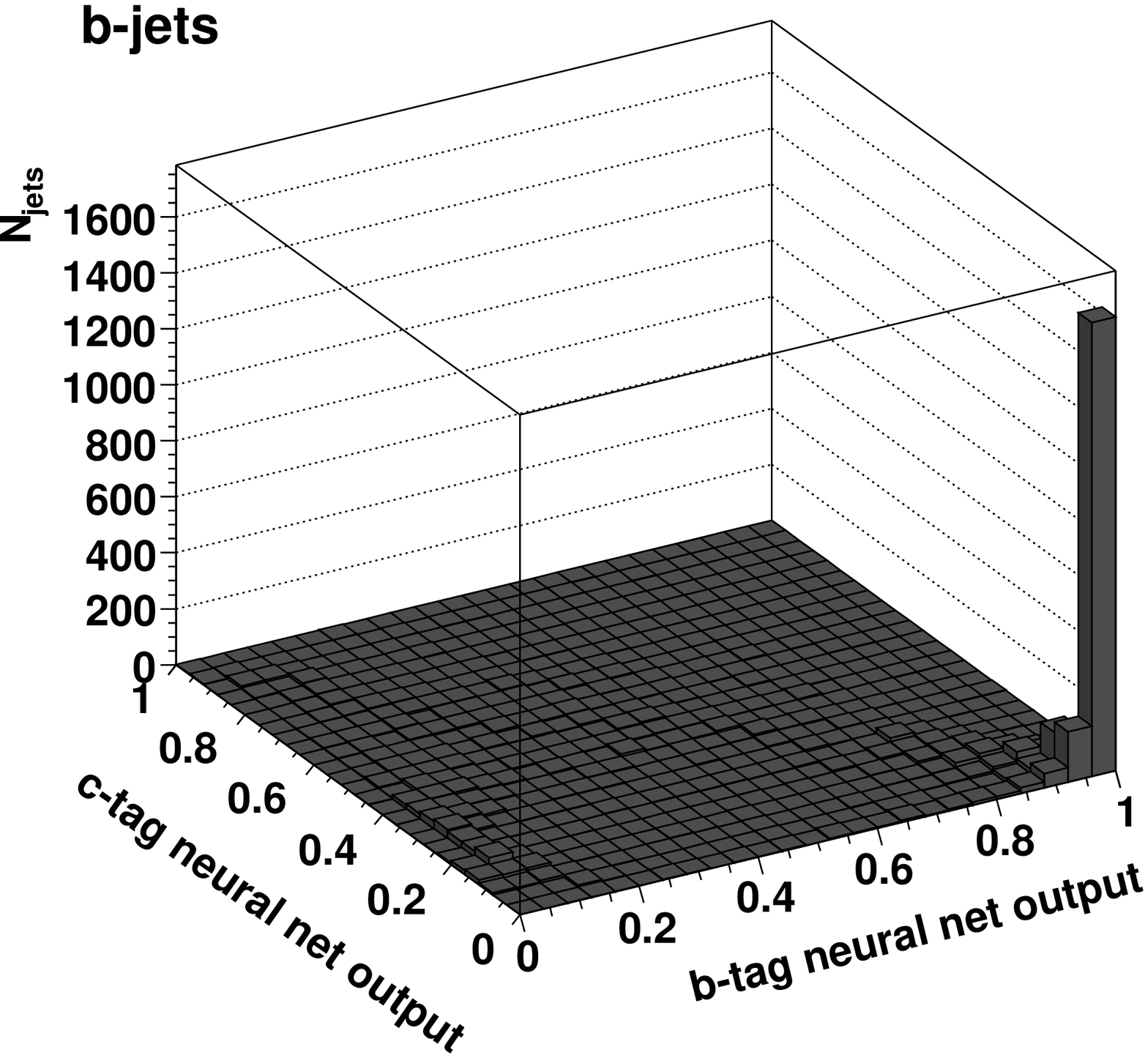} &
\includegraphics[width=0.50\columnwidth]{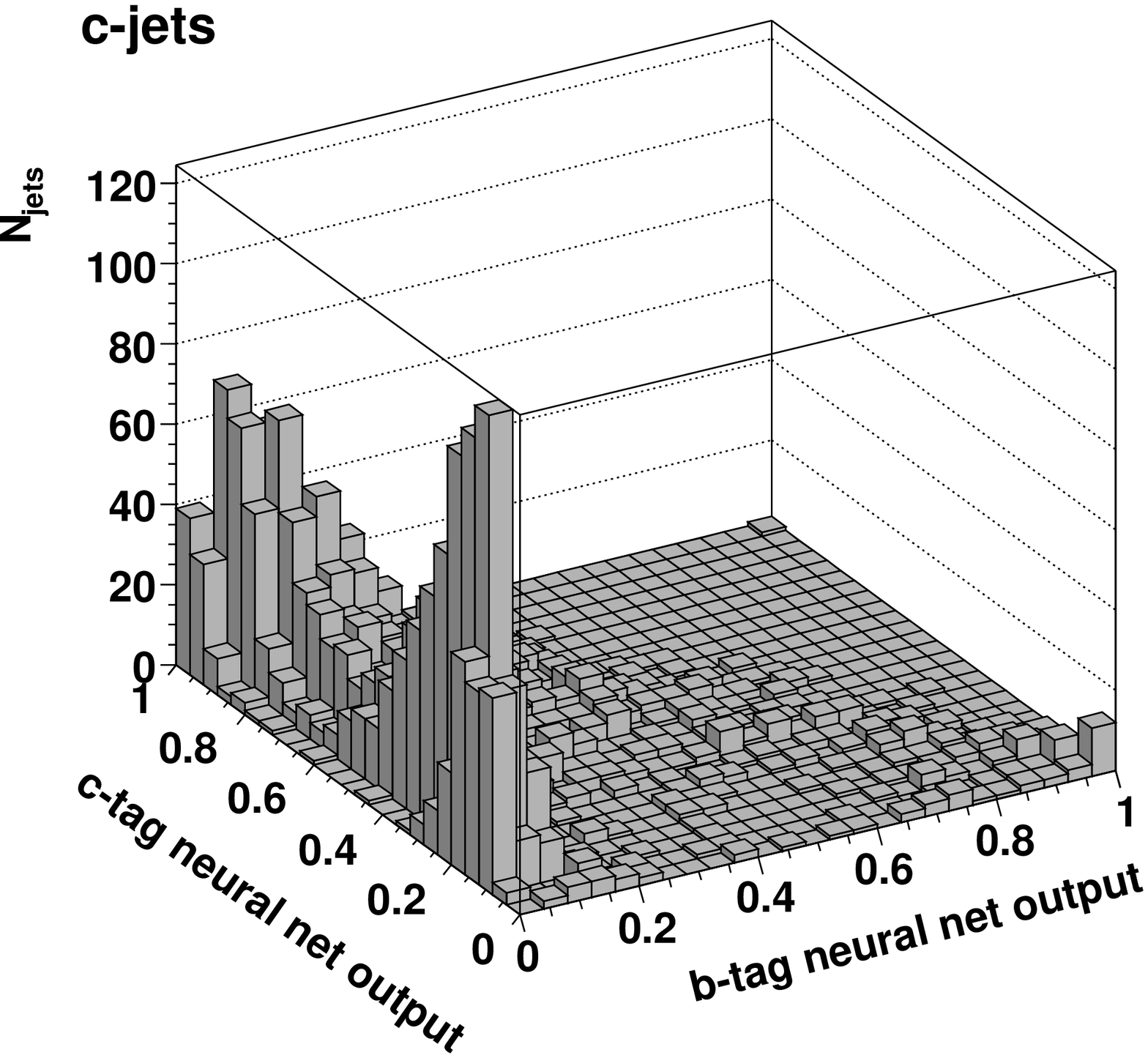}\\[-35ex]
\hspace*{12em}(a) & \hspace*{12em}(b) \\[30ex]
\includegraphics[width=0.50\columnwidth]{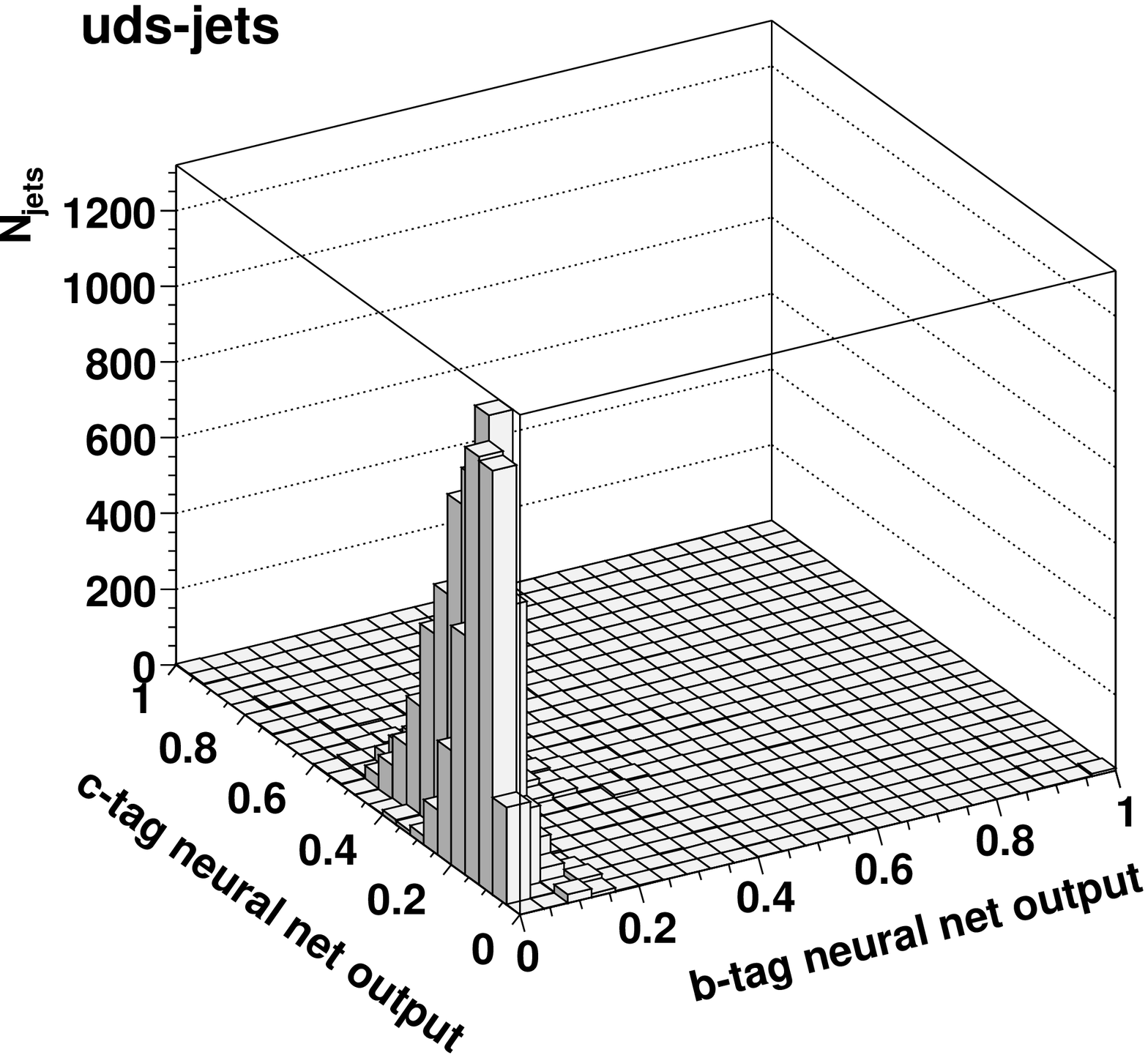} & \\[-35ex]
\hspace*{12em}(c) & \\[30ex]
\end{tabular}
\caption{\textit{Charm tag vs.~bottom tag for input samples consisting
purely of (a) bottom jets, (b) charm jets and (c) light quark jets.}}
\label{FigureCTagVsBTag}
\end{center}
\end{figure}
Figure~\ref{FigureCTagOutput} shows the distribution of the output 
variables of neural networks used for tagging charm jets ($c$ nets) 
separately for the cases of one, two and at least three vertices, 
and the combined distribution for an arbitrary number of vertices, 
for the sample of two-jet events at $\sqrt{s} = 91.2\,\mathrm{GeV}$.
The most straightforward way of using the charm tag (i.e.~$c$ net
output) in an analysis  is to require one or more jets in an event to
have a charm tag exceeding a certain cut value, chosen as appropriate
for the specific analysis.
Resulting performance on a jet-by-jet basis is discussed in Section 
\ref{SubsectionFlavourTagPerformance}.
Event selection can be improved by using information from both the
charm and the bottom tag.
This can, for example, be achieved by plotting charm versus 
bottom tag, as shown in Fig.~\ref{FigureCTagVsBTag} for bottom, charm 
and light flavour jets from the two-jet $Z$ peak sample, and placing 
a cut on the resulting two-dimensional distribution.
Note that in the two-dimensional distribution for $c$ jets, the 
peak near ($b$-tag $= 0$, $c$-tag $= 0$) stems from jets in which 
only the primary vertex was reconstructed, while the peak near
$(0,1)$ is due to jets in which secondary vertices were also 
found.

\subsection{Functionality provided by the LCFIVertex neural net code}
\label{SubsectionNeuralNetCode}

The flavour tag described in the previous section is based on a 
neural network approach.
Within the LCFIVertex package, neural network code implementing
flexible multi-layer perceptrons is provided which was originally
developed as a standalone package.
In addition to the flavour tag processor already described, a
dedicated Marlin processor is provided to train new networks.
This can, for example, be used for training dedicated networks
for specific analyses or detector geometries (although as a 
general rule the networks trained with two-jet events provide
excellent tagging performance and are very widely usable).
Furthermore, this processor provides an example of how a 
neural network can be set up and trained, for users who wish
to change the flavour tagging approach, e.g.~by using further
input variables, different network architecture and/or training
algorithms, or who would like to set up networks for new
purposes such as the tagging of $\tau$ leptons, currently not
available in the LCFIVertex code.
This section gives an overview of the functionality of the 
generic neural network code provided.

In a multi-layer perceptron, the value $a_{i}$ that is passed to  
the transfer function of neuron $i$ is obtained from the weighted 
sum of the outputs $t_j$ from all neurons in the preceding layer, 
with the option to subtract a bias value $w_{ib}$:
$a_{i} = \sum_{j=1}^{N} t_{j} w_{ij} - w_{ib}$.
Four training algorithms to adjust the weights of the network are 
implemented, with the initial weights set to random values.
The four training algorithms are:
\begin{enumerate}
\item{the back propagation algorithm;}
\item{the back batch propagation algorithm;}
\item{the back propagation conjugate gradient (CG) algorithm;}
\item{the genetic algorithm.}
\end{enumerate}
These are described in more detail in the literature on neural 
nets, for introductions to the subject see 
e.g.~Refs.~\cite{Bishop:1995,Hertz:Krogh:Palmer:1991}.
Algorithms 1 and 2 only differ in the number of training items
processed per iteration; algorithm 3 uses a conjugate gradient
approach to minimise the error function by optimal choice of 
the two parameters (learning rate and momentum constant) that
influence the weight changes at each training step.
This algorithm converges considerably faster, by a factor of
order 10, than the simple back propagation algorithm and is 
used as the default training algorithm in the LCFIVertex
package.
Algorithm 4 is not based on error function minimisation, but 
during the training phase works with a set of neural networks,
with the probability of one or more copies of a network being
retained from one training step to the next being the larger
the better a network performs on the training sample.

\subsection{Performance of the flavour tag}
\label{SubsectionFlavourTagPerformance}

\begin{figure}[htb]
\begin{center}
\includegraphics[width=0.95\columnwidth]{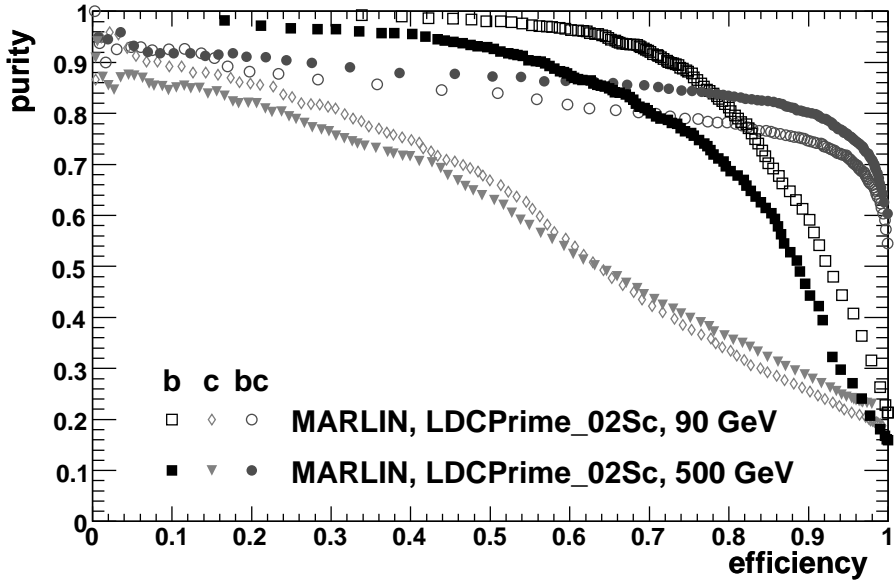}
\caption{\textit{Comparison of tagging performance at the $Z$ resonance and at 
$\sqrt{s} = 500\,\mathrm{GeV}$. Tagging purity is plotted as function of 
efficiency for $b$ jets and $c$ jets. Performance for $c$ jets assuming only
$b$ background (labelled ``bc'') is also shown.}}
\label{FigureFlavourTagPerformance}
\end{center}
\end{figure}
As a measure of the flavour-tagging performance, the purity of
selecting bottom and charm jets is studied as function of the
efficiency.
Fig.~\ref{FigureFlavourTagPerformance} shows purity vs.~efficiency 
for the three tags provided by the LCFIVertex package, for the 
two-jet sample at $\sqrt{s} = 91.2\,\mathrm{GeV}$ and at 
$\sqrt{s} = 500\,\mathrm{GeV}$.
A flavour composition of approximately $22\,\%$ 
($15\,\%$) of bottom, $17\,\%$ ($25\,\%$) of charm and about
$61\,\%$ ($60\,\%$) of light flavour jets at 
$\sqrt{s} = 91.2\,\mathrm{GeV}$ ($500\,\mathrm{GeV}$) is
assumed.
The plot is obtained by varying a simple cut on the neural 
net output variable and calculating purity and efficiency at
each cut value.
At the $Z$ resonance, for the $b$-tag a very pure sample, 
containing $92\,\%$ $b$ jets, can be selected at an efficiency
of $70\,\%$.
In comparison, high $c$-tag purities can only be achieved at lower
efficiencies, mainly
{due to contamination from light flavour jets.}
The $c$-tag with all backgrounds included has been found to 
be the most sensitive of the tags whenever changing boundary
conditions for the study, such as using different tracking
algorithms, code parameters and detector geometry.
At $500\,\mathrm{GeV}$, the $b$-tag performance is degraded
with respect to that at the $Z$ resonance, while $c$-tag
purity is very similar at both energies and the purity of
the $c$-tag with only $b$ background is slightly improved
at the higher energy.

\begin{figure}[htb]
\begin{center}
\begin{tabular}{cc}
\includegraphics[width=0.50\columnwidth]{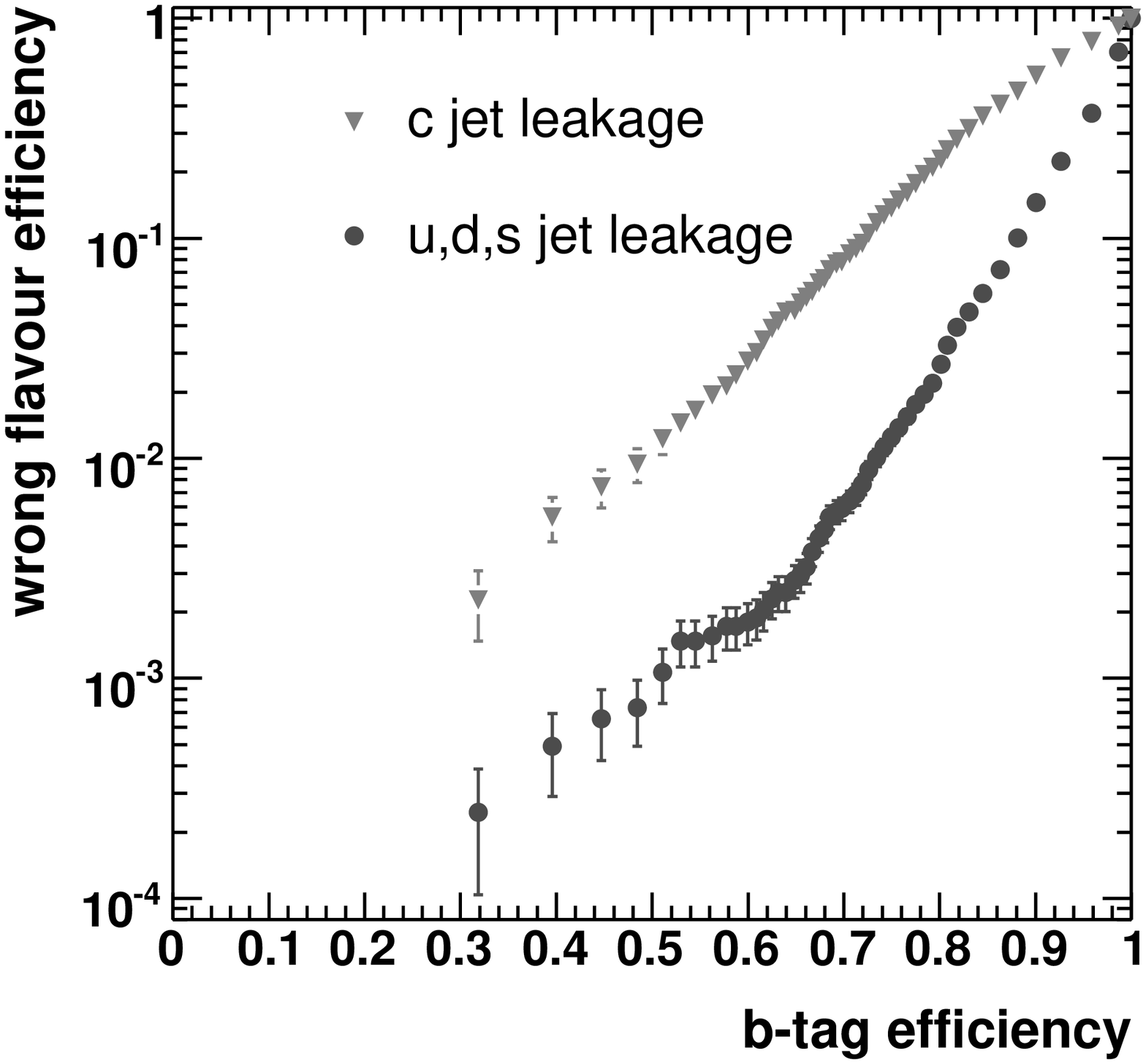} &
\includegraphics[width=0.50\columnwidth]{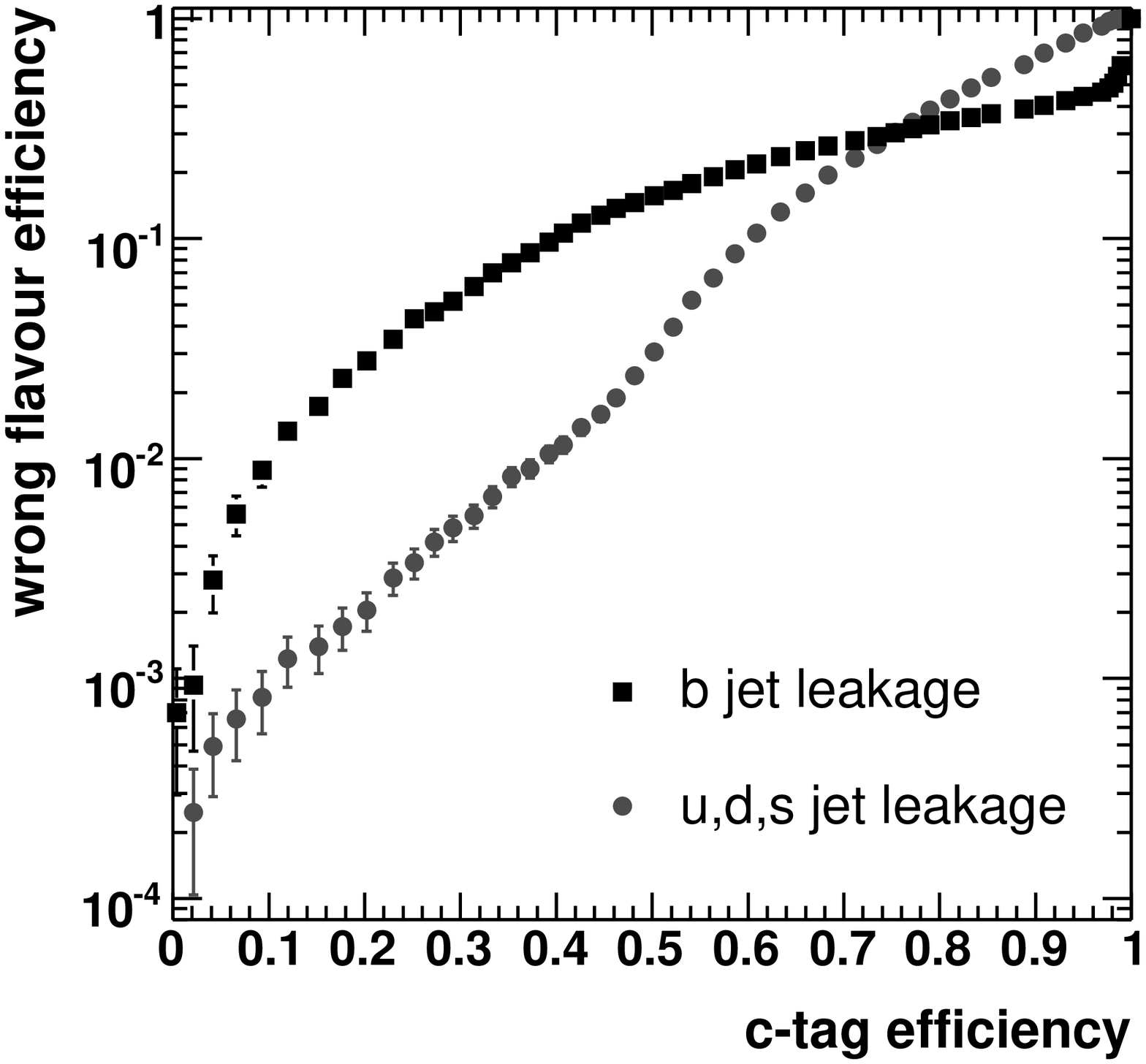}\\[-35ex]
\hspace*{-9em}(a) & \hspace*{-9em}(b) \\[30ex]
\end{tabular}
\caption{\textit{Efficiencies for selecting jets with the wrong flavour
when tagging (a) bottom jets and (b) charm jets.}}
\label{FigureFlavourLeakageRates}
\end{center}
\end{figure}
A way of studying tagging performance that is independent of 
the sample composition is to look at the efficiency of 
selecting each of the ``wrong'' flavours when cutting on one
of the tagging variables, i.e.~the wrong flavour efficiencies.
Figure~\ref{FigureFlavourLeakageRates} shows the $c$ and light 
quark jet efficiencies as a function of the $b$-tag efficiency 
and the $b$ and light flavour jet efficiencies vs.~the $c$-tag
efficiency.
It is clearly seen that the main background to the $b$-tag is
due to $c$ jets.
For the $c$-tag the main background at efficiencies above
about $75\,\%$ is due to light flavour jets, while at lower
efficiencies it is dominated by misidentified $b$ jets, as
would be expected from the comparison of the purities for
$c$-tag and $c$-tag with $b$ background only.

\begin{table}[hp]
\begin{center}
\begin{tabular}{||c||c|c|c||}
\hline
\multicolumn{4}{||c||}{1 vertex found by ZVTOP} \\ \hline
variable & $b$-tag & $c$-tag & $c$-tag ($b$ bgd.) \\ 
\hline
most signif.~track, $d_{0}/\sigma(d_{0})$                   & $0.002 \pm 0.004$ & $0.005 \pm 0.008$ & $0.000 \pm 0.000$ \\
$2^{\textrm{nd}}$-most signif.~track, $d_{0}/\sigma(d_{0})$ & $0.058 \pm 0.016$ & $0.009 \pm 0.007$ & $0.026 \pm 0.010$ \\
most signif.~track, $z_{0}/\sigma(z_{0})$                   & $0.001 \pm 0.001$ & $0.003 \pm 0.002$ & $0.002 \pm 0.001$ \\
$2^{\textrm{nd}}$-most signif.~track, $z_{0}/\sigma(z_{0})$ & $0.115 \pm 0.030$ & $0.005 \pm 0.001$ & $0.066 \pm 0.017$ \\
most signif.~track, $|p_{\mathrm{trk}}|$                    & $0.113 \pm 0.005$ & $0.275 \pm 0.015$ & $0.043 \pm 0.002$ \\
$2^{\textrm{nd}}$-most signif.~track, $|p_{\mathrm{trk}}|$  & $0.062 \pm 0.003$ & $0.068 \pm 0.006$ & $0.048 \pm 0.003$ \\
joint probability, $R$-$\phi$                               & $1$ & $0.676 \pm 0.031$ & $0.887 \pm 0.041$ \\
joint probability, $R$-$z$                                  & $0.922 \pm 0.037$ & $1$ & $1$ \\
\hline\hline
\multicolumn{4}{||c||}{2 vertices found by ZVTOP} \\ \hline
variable & $b$-tag & $c$-tag & $c$-tag ($b$ bgd.) \\ 
\hline
$M_{\mathrm{Pt}}$                   & $1$ & $1$ & $1$ \\
$|p|$                               & $0.098 \pm 0.007$ & $0.404 \pm 0.029$ & $0.114 \pm 0.008$ \\
decay length                        & $0.182 \pm 0.013$ & $0.990 \pm 0.072$ & $0.139 \pm 0.010$ \\
decay length significance           & $0.070 \pm 0.008$ & $0.187 \pm 0.022$ & $0.115 \pm 0.013$ \\
$N_{\mathrm{trk,vtx}}$              & $0.063 \pm 0.004$ & $0.162 \pm 0.009$ & $0.081 \pm 0.004$ \\
secondary vertex probability        & $0.230 \pm 0.017$ & $0.212 \pm 0.016$ & $0.124 \pm 0.010$ \\
joint probability, $R$-$\phi$       & $0.040 \pm 0.004$ & $0.071 \pm 0.007$ & $0.052 \pm 0.005$ \\
joint probability, $R$-$z$          & $0.061 \pm 0.008$ & $0.159 \pm 0.021$ & $0.052 \pm 0.007$ \\
\hline\hline
\multicolumn{4}{||c||}{3 vertices found by ZVTOP} \\ \hline
variable & $b$-tag & $c$-tag & $c$-tag ($b$ bgd.) \\ 
\hline
$M_{\mathrm{Pt}}$                   & $1$ & $1$ & $1$ \\
$|p|$                               & $0.490 \pm 0.076$ & $0.960 \pm 0.148$ & $0.580 \pm 0.089$ \\
decay length                        & $0.550 \pm 0.119$ & $0.338 \pm 0.072$ & $0.370 \pm 0.080$ \\
decay length significance           & $0.081 \pm 0.018$ & $0.106 \pm 0.023$ & $0.078 \pm 0.017$ \\
$N_{\mathrm{trk,vtx}}$              & $0.665 \pm 0.089$ & $0.811 \pm 0.108$ & $0.829 \pm 0.111$ \\
secondary vertex probability        & $0.087 \pm 0.045$ & $0.112 \pm 0.057$ & $0.066 \pm 0.034$ \\
joint probability, $R$-$\phi$       & $0.006 \pm 0.002$ & $0.012 \pm 0.005$ & $0.007 \pm 0.003$ \\
joint probability, $R$-$z$          & $0.007 \pm 0.003$ & $0.017 \pm 0.007$ & $0.009 \pm 0.003$ \\
\hline
\end{tabular}
\vspace*{2ex}
\caption{\textit{Relative importance $I_{i}/I_{\mathrm{max}}$ of variables used as inputs for
flavour tag neural nets at the $Z$ resonance.}}
\label{TableFlavourTagInputsImportance91GeV}
\end{center}
\end{table}
%
%
\begin{table}[hp]
\begin{center}
\begin{tabular}{||c||c|c|c||}
\hline
\multicolumn{4}{||c||}{1 vertex found by ZVTOP} \\ \hline
variable & $b$-tag & $c$-tag & $c$-tag ($b$ bgd.) \\ 
\hline
most signif.~track, $d_{0}/\sigma(d_{0})$                   & $0.363 \pm 0.084$ & $1$ & $0.063 \pm 0.014$ \\
$2^{\textrm{nd}}$-most signif.~track, $d_{0}/\sigma(d_{0})$ & $0.499 \pm 0.079$ & $0.072 \pm 0.020$ & $0.429 \pm 0.067$ \\
most signif.~track, $z_{0}/\sigma(z_{0})$                   & $0.036 \pm 0.008$ & $0.166 \pm 0.053$ & $0.148 \pm 0.033$ \\
$2^{\textrm{nd}}$-most signif.~track, $z_{0}/\sigma(z_{0})$ & $1$ & $0.057 \pm 0.016$ & $1$ \\
most signif.~track, $|p_{\mathrm{trk}}|$                    & $0.019 \pm 0.004$ & $0.057 \pm 0.019$ & $0.013 \pm 0.003$ \\
$2^{\textrm{nd}}$-most signif.~track, $|p_{\mathrm{trk}}|$  & $0.009 \pm 0.002$ & $0.013 \pm 0.005$ & $0.012 \pm 0.003$ \\
joint probability, $R$-$\phi$                               & $0.364 \pm 0.066$ & $0.304 \pm 0.091$ & $0.564 \pm 0.102$ \\
joint probability, $R$-$z$                                  & $0.363 \pm 0.061$ & $0.487 \pm 0.142$ & $0.688 \pm 0.116$ \\
\hline\hline
\multicolumn{4}{||c||}{2 vertices found by ZVTOP} \\ \hline
variable & $b$-tag & $c$-tag & $c$-tag ($b$ bgd.) \\ 
\hline
$M_{\mathrm{Pt}}$                   & $1$ & $0.438 \pm 0.116$ & $1$ \\
$|p|$                               & $0.071 \pm 0.021$ & $0.128 \pm 0.034$ & $0.082 \pm 0.024$ \\
decay length                        & $0.420 \pm 0.119$ & $1$ & $0.320 \pm 0.091$ \\
decay length significance           & $0.011 \pm 0.006$ & $0.013 \pm 0.007$ & $0.019 \pm 0.011$ \\
$N_{\mathrm{trk,vtx}}$              & $0.054 \pm 0.014$ & $0.061 \pm 0.013$ & $0.070 \pm 0.018$ \\
secondary vertex probability        & $0.262 \pm 0.078$ & $0.106 \pm 0.028$ & $0.141 \pm 0.042$ \\
joint probability, $R$-$\phi$       & $0.083 \pm 0.026$ & $0.064 \pm 0.018$ & $0.108 \pm 0.034$ \\
joint probability, $R$-$z$          & $0.147 \pm 0.057$ & $0.168 \pm 0.061$ & $0.125 \pm 0.049$ \\
\hline\hline
\multicolumn{4}{||c||}{3 vertices found by ZVTOP} \\ \hline
variable & $b$-tag & $c$-tag & $c$-tag ($b$ bgd.) \\ 
\hline
$M_{\mathrm{Pt}}$                   & $0.500 \pm 0.268$ & $0.814 \pm 0.435$ & $0.745 \pm 0.398$ \\
$|p|$                               & $0.173 \pm 0.096$ & $0.553 \pm 0.307$ & $0.306 \pm 0.169$ \\
decay length                        & $1$ & $1$ & $1$ \\
decay length significance           & $0.013 \pm 0.010$ & $0.028 \pm 0.021$ & $0.019 \pm 0.014$ \\
$N_{\mathrm{trk,vtx}}$              & $0.418 \pm 0.172$ & $0.829 \pm 0.341$ & $0.775 \pm 0.318$ \\
secondary vertex probability        & $0.349 \pm 0.504$ & $0.731 \pm 1.050$ & $0.394 \pm 0.569$ \\
joint probability, $R$-$\phi$       & $0.015 \pm 0.011$ & $0.048 \pm 0.036$ & $0.028 \pm 0.021$ \\
joint probability, $R$-$z$          & $0.035 \pm 0.033$ & $0.127 \pm 0.123$ & $0.062 \pm 0.061$ \\
\hline
\end{tabular}
\vspace*{2ex}
\caption{\textit{Relative importance $I_{i}/I_{\mathrm{max}}$ of variables used as inputs for
flavour tag neural nets at $500\,\mathrm{GeV}$.}}
\label{TableFlavourTagInputsImportance500GeV}
\end{center}
\end{table}
The relative importance of the input variables on the flavour
tag results for the different contributing neural nets was
determined using an estimator also used in the ``Toolkit for
Multivariate Data Analysis with ROOT'' (TMVA) \cite{TMVA:2007}.
Following this approach, the input importance of variable
$i$ is defined as
\[ I_{i} = \bar{x}_{i}^{2} \sum_{j=1}^{n_{h}} 
         \left( w_{ij}\right)^{2}\ \ \ i=1,...,n_{\mathrm{var}},
\]
where $\bar{x}_{i}$ is the average of the values of variable
$i$ in the input sample and the sum extends over the weights $w_{ij}$
corresponding to the connections of the neural network node
of variable $i$ with the $n_{h}$ nodes in the adjacent network
layer.
The calculation was implemented as part of the neural network
code provided with LCFIVertex and this information is provided
by the flavour tag processor as part of the run summary.
Tables \ref{TableFlavourTagInputsImportance91GeV} and 
\ref{TableFlavourTagInputsImportance500GeV} summarise the 
results obtained at the $Z$ resonance and at 
$500\,\mathrm{GeV}$, respectively, where for each neural
network, values $I_{i}$ are normalised to the maximum value
$I_{\mathrm{max}}$.

At the $Z$ resonance, in case only the primary vertex is found
by ZVTOP, the joint probability variables in $R$-$\phi$ and $z$
provide the best handle for distinguishing between different
jet flavours.
This is to be expected given that these variables combine
information from all the tracks in the jet, rather than 
resulting from only one of them (as is the case for the other
six inputs).
For the $c$-tag provided for the case that all backgrounds are
present, the momentum of the most significant track in the jet
also contributes significantly to the flavour tag result.
The other variables contribute to a much lesser extent.
At $500\,\mathrm{GeV}$, the joint probability variables are 
still important, but the impact parameter significances of the
most and second most significant track in the jet play a
similarly important role in jet flavour identification.

If at least two vertices are found in a jet, the 
$P_{t}$-corrected vertex mass provides the clearest indication
of the jet flavour.
Other important variables are the seed vertex decay length, 
particularly if exactly two vertices are found, and the number
of tracks in the seed vertex, especially if three or more
vertices are found, as well as the vertex momentum $|p|$ and
the secondary vertex probability.
As can be expected, the relative importance of the decay length
increases with increasing jet energy, and surpasses that of 
the $M_{\mathrm{Pt}}$ variable for the three-vertex case at
$500\,\mathrm{GeV}$.

\section{Quark charge determination}
\label{SectionQuarkCharge}

\subsection{Quark charge for $b$ and $c$ jets}

Quark charge determination, i.e.~reconstruction of the charge
sign of the heaviest quark in the leading hadron of a jet, 
will be {important} for a range of measurements at a Linear 
Collider.
An example is the measurement of the left-right asymmetry
$A_{\mathrm{LR}}$ in 
$e^{+}e^{-} \rightarrow \gamma /Z \rightarrow b\bar{b}$ and
$e^{+}e^{-} \rightarrow \gamma /Z \rightarrow c\bar{c}$, which 
is sensitive to new physics phenomena beyond the direct energy 
reach of the ILC and LHC, such as $Z'$ exchange, leptoquarks, 
$R$-parity-violating scalar particles and extra spatial 
dimensions \cite{Hewett:1999,Riemann:2001}.
Quark charge measurement can also help reduce combinatoric
backgrounds in multi-jet events.

The algorithm for reconstructing the quark charge differs for
$b$ and $c$ jets.
Flavour tagging is therefore a prerequisite for quark charge
determination.
Reconstruction is more challenging for $b$ than for $c$ jets,
as $b$ decay chains tend to be more complex than $c$ decays.
In a pure $b$ jet sample, the $b$ hadron is charged in about
$40\,\%$ of the jets.
For the cases in which a non-zero vertex charge is found, the
quark charge sign is given directly by the vertex charge.

To calculate the vertex charge, all tracks that are thought to
belong to the decay chain are identified and the vertex charge
is given by the sign of the sum of their charges.
The rare cases in which the vertex charge has an absolute 
value above 2, indicating imperfections in the assignment of 
tracks to the decay chain, are discarded in the determination
of the quark charge sign.
To assign tracks to the decay chain, initially the same 
approach is used as is described in Section 
\ref{SubsectionFlavourTagInputs} for $M_{\mathrm{Pt}}$.
In addition, if more than two vertices are found, the
tracks contained in the secondary vertex are included in
the $Q_{\mathrm{sum}}$ calculation, even if they fail the
$L/D$ cut, which helps in the case of charm decays with a
long decay length, corresponding to a large distance 
$D$ of the charm decay from the IP and hence a 
comparatively small $L/D$ value for the tracks from the
$B$ decay vertex.

For $c$ jets, the vertex charge is given by the charge
sum, $Q_{\mathrm{sum}}$, in the same way as for $b$ jets,
with the same algorithm being used for the assignment of
tracks to the decay chain as in the calculation of
$M_{\mathrm{Pt}}$.
The special procedure for the assignment of tracks from a
secondary vertex as described above for $b$ jets with more
than two vertices is not applied, as there is no motivation
for it for $c$ jets, and it was found to degrade performance.

{
\begin{figure}[htb]
\begin{center}
\begin{tabular}{cc}
\includegraphics[width=0.50\columnwidth]{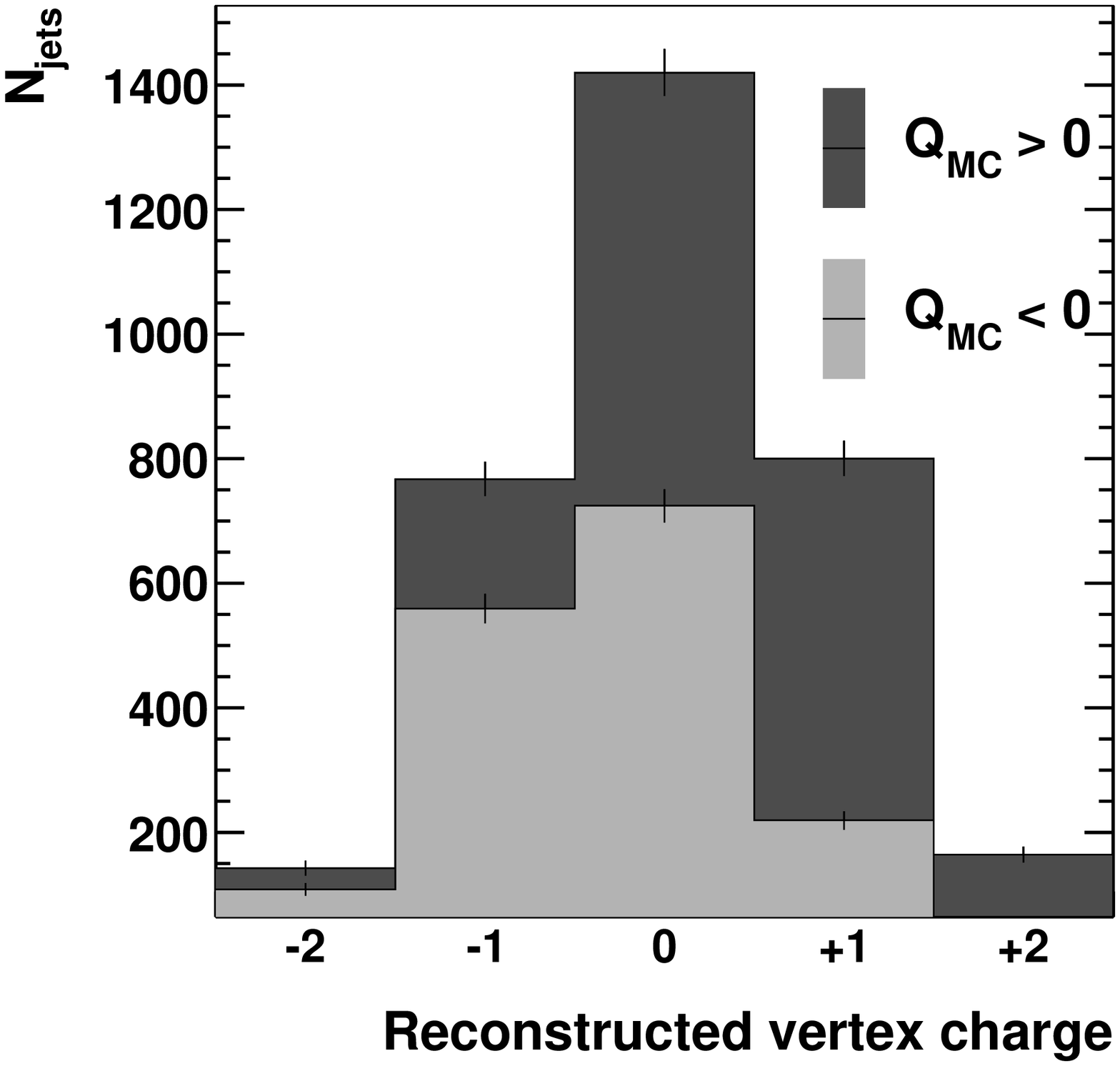} &
\includegraphics[width=0.50\columnwidth]{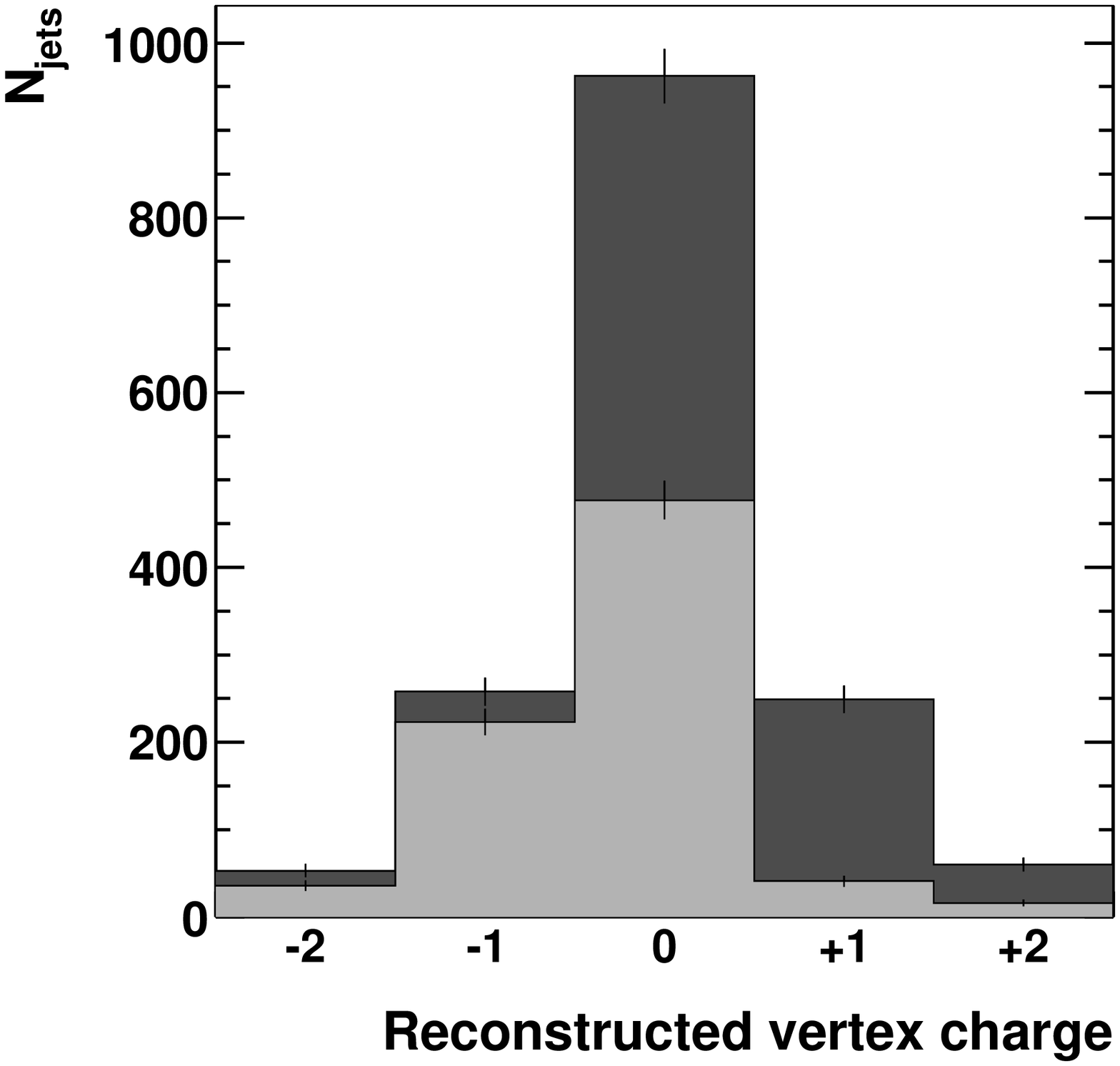}\\[-35ex]
\hspace*{-7em}(a) & \hspace*{-7em}(b) \\[30ex]
\end{tabular}
\caption{\textit{Reconstructed vertex charge in comparison to the parton charge
in (a) bottom jets and (b) charm jets, at the $Z$ resonance.}}
\label{FigureVertexCharge}
\end{center}
\end{figure}
\begin{table}[htb]
\begin{center}
\begin{tabular}{||cc|cccc||}
\hline
\multicolumn{2}{||c|}{Monte Carlo} &
\multicolumn{4}{c||}{Reconstructed vertex charge} \\
\multicolumn{2}{||c|}{jet type} &
\multicolumn{4}{c||}{ } \\
\hline
                     &            & correct & ambiguous & wrong & no tracks \\
\hline
$91.2\,\mathrm{GeV}$ & $b$ jets   &  32.8   &   33.1    & 12.2  &  21.9     \\
                     & $c$ jets   &  14.6   &   27.4    &  3.2  &  54.8     \\
\hline
$500\,\mathrm{GeV}$  & $b$ jets   &  27.0   &   23.9    & 13.5  &  35.6     \\
                     & $c$ jets   &  13.0   &   24.5    &  4.2  &  58.4     \\
\hline
\end{tabular}
%
%
%
%
%
%
%
\vspace*{2ex}
\caption{\textit{Performance of parton charge identification using LCFIVertex
  default settings. Non-zero vertex charge measurement with the same (opposite)
  sign as the parton charge is labelled ``correct'' (``wrong''). Jets with
  a reconstructed vertex charge of zero are called ``ambiguous''. Jets where
  no tracks pass the strict track quality criteria for the vertex charge
  measurement are listed in the rightmost column.
  }}
\label{TableVertexCharge}
\end{center}
\end{table}
Figure~\ref{FigureVertexCharge} demonstrates the ability to
distinguish the parton charge sign in heavy flavour jets. No
information about the parton charge is obtained for the case of a
reconstructed vertex charge of zero, the dominant contribution to
which is from vertices with exactly two tracks attached. Good charge
separation can be achieved by selecting jets with non-zero
reconstructed vertex charge. Table~\ref{TableVertexCharge} quantifies the
performance of parton charge identification with standard LCFIVertex
parameters. The ratio of correct over wrong charge identification is
better for $c$ jets than for $b$ jets, and the reconstruction quality is
found to degrade slightly with increasing collision energy as expected.
}

\section{Summary}
\label{SectionSummary}

Precision measurements at the International Linear Collider
will rely on excellent vertexing capabilities of the vertex
detector and the reconstruction software.
The LCFIVertex software provides vertexing, flavour tagging
and vertex charge reconstruction algorithms which are being
used for optimisation of the ILC detectors in the current 
R\&D phase of the project.

Two vertexing algorithms are provided, which were originally
developed at the SLD experiment: the ZVRES algorithm, being
more generally applicable, and the ZVKIN algorithm, being
specifically tailored to vertex finding for bottom jets.
For a typical ILC detector design and using two-jet events at
$\sqrt{s} = 91.2\,\mathrm{GeV}$, the ZVRES approach is 
expected to provide a secondary vertex finding efficiency 
of $\approx 90\,\%$ for $b$ jets.
At $500\,\mathrm{GeV}$, the assignment of tracks to 
vertices is affected by the increased level of 
final-state radiation and the decreased opening angles of
jets.
The LCFIVertex implementation of ZVRES takes
the energy dependence into account in the definition of 
the vertex function, the mathematical representation of 
the jet topology.
A dedicated study of how vertexing performance varies 
with jet energy could result in further improvements.

The flavour tagging algorithm provided by the LCFIVertex
package is based on the use of artificial neural networks.
Different networks are used for bottom and charm jets and
for the cases that one, two or at least three vertices
were found in the input jet.
For two-jet events at the $Z$ resonance, $b$ jets can be
selected with $\approx 90\,\%$ purity for an efficiency
of $70\,\%$.
At $500\,\mathrm{GeV}$, the corresponding $b$-tag 
purity is $\approx 80\,\%$.
A method for monitoring the relative importance of the
flavour tag input variables for each of the neural 
networks is provided with
the code.
Important input variables include vertex mass, momentum
and decay length as well as the multiplicity of tracks
assigned to vertices.
Opportunities for future studies in this area include
the change of parameters used in the calculation of the
current flavour tag inputs, the addition of further
variables, change of neural network architecture and
exploration of other classification approaches such
as boosted decision trees.

\appendix

\section{Joint probability: parameters used in the calculation}
\label{AppendixJointProbability}

It follows from the definition of the joint probability $P_{J}$, see
Section \ref{SubsectionFlavourTagInputs}, that it depends on the 
distribution of impact parameter significances of IP tracks.
The function $f(x)$ that approximates this distribution is determined
from the data as follows:
for IP tracks, the impact parameter significance distributions are
symmetric.
The shape of the distribution of positive impact parameters, used to 
find $P_{J}$, can therefore be determined by fitting the distribution
of absolute values for tracks with negative impact parameter, 
corresponding to a very pure IP track sample.
The distribution is approximated by fitting the sum of a Gaussian and
two exponentials to the distribution, i.e.~by finding parameters
$p_{f0}$, $\ldots$, $p_{f6}$ such that
\[ f(x) =   p_{f0} \cdot 
                    \exp{ \left( -0.5 \left( \frac{x-p_{f1}}{p_{f2}}\right)^{2} \right) }
          + \exp{\left( p_{f3} + p_{f4} x \right)}
	  + \exp{\left( p_{f5} + p_{f6} x \right)}
\]
describes the measured distribution.
With this approximation, the integral $\int_{b/\sigma_{b}}^{\infty}{f(x)dx}$ 
can be written in the form
\begin{align*}
  \int_{b/\sigma_{b}}^{\infty} f(x) dx \approx & {} 
    \frac{2}{\sqrt{\pi}} \cdot
    \left(
      \int_{(b/\sigma_{b})/\left( \sqrt{2} p_{I0} \right)}^{\infty} \exp{\left( -r^{2} \right)} dr -
      \int_{(b/\sigma_{b})_{\mathrm{cut}}/\left( \sqrt{2} p_{I0} \right)}^{\infty} \exp{\left( -r^{2} \right)} dr
    \right) \\[1ex]
    & + p_{I1} \left( \exp{\left(-p_{I2}x\right)} -
                        \exp{\left(-p_{I2}b/\sigma_{b}\right)} \right) \\[1ex]
    & + p_{I3} \left( \exp{\left(-p_{I4}x\right)} -
                        \exp{\left(-p_{I4}b/\sigma_{b}\right)} \right) \ \ \ , \\
\end{align*}
where the integral over the Gaussian is cut off at $(b/\sigma_{b})_{\mathrm{cut}}$ 
and the following parameter transformation is used:
\begin{align*}
p_{I0} = {} & p_{f2} \\
p_{I1} = {} & - \frac{2}{\sqrt{\pi}} \exp{\left( p_{f3}/ \left( p_{f0}p_{f2}p_{f4} \right)  \right)} \\
p_{I2} = {} & - p_{f4} \\
p_{I3} = {} & - \frac{2}{\sqrt{\pi}} \exp{\left( p_{f5}/ \left( p_{f0}p_{f2}p_{f6} \right)  \right)} \\
p_{I4} = {} & - p_{f6} \\
\end{align*}
\begin{table}[htb]
\begin{center}
\begin{tabular}{||c|c|c||}
\hline
parameter & joint probability in $R$-$\phi$ & joint probability in $z$ \\
\hline
$p_{I0}$ & 0.843 & 0.911 \\
$p_{I1}$ & 0.365 & 0.306 \\
$p_{I2}$ & 0.620 & 0.423 \\
$p_{I3}$ & 0.150 & 0.139 \\
$p_{I4}$ & 0.029 & 0.028 \\
\hline
\end{tabular}
\vspace*{2ex}
\caption{\textit{Parameters used in the calculation of the joint probability in
$R$-$\phi$ and in $z$, respectively, obtained from a fit to negative impact
parameter distributions.}}
\label{TableJointProbabilityParameters}
\end{center}
\end{table}
In the code, the fit and parameter transformation are performed in the
same processor.
For the detector geometry \verb|LDCPrime_02Sc| it yields the parameters
listed in Table \ref{TableJointProbabilityParameters}.

\section{Values of code parameters}
\label{AppendixCodeParameters}

\begin{table}[hbt]
\begin{center}
\begin{tabular}{||c|c||c|c||c|c||}
\hline
\multicolumn{2}{||c||}{ZVRES} &
\multicolumn{2}{c||}{flavour tag} &
\multicolumn{2}{c||}{vertex charge} \\ \hline
parameter & value & parameter & value & parameter & value \\ 
\hline
$w_{\mathrm{IP}}$          & 1     &  $N_{L}$                                     & 5    & $T_{\mathrm{qb,max}}$ ($\mathrm{mm}$) & 1    \\ \hline
$k$                        & 0.125 &  $p_{\mathrm{trk,NL,min}}$ ($\mathrm{GeV}$)  & 1    & $\left(L/D\right)_{\mathrm{qb,max}}$  & 2.5  \\ \hline
$R_{0}$                    & 0.6   &  $p_{\mathrm{trk,NL-1,min}}$($\mathrm{GeV}$) & 2    & $\left(L/D\right)_{\mathrm{qb,min}}$  & 0.18 \\ \hline
$\chi^{2}_{\mathrm{TRIM}}$ & 10    &  $N_{\mathrm{trks,min}}$                     & 1    & $T_{\mathrm{qc,max}}$ ($\mathrm{mm}$) & 1    \\ \hline
$\chi_{0}^{2}$             & 10    &  $\chi_{\mathrm{norm,max}}^{2}$              & 20   & $\left(L/D\right)_{\mathrm{qc,max}}$  & 2.5  \\ \hline
                           &       &  $T_{\mathrm{max}}$ ($\mathrm{mm}$)          & 1    & $\left(L/D\right)_{\mathrm{qc,min}}$  & 0.5  \\ \hline
                           &       &  $\left(L/D\right)_{\mathrm{max}}$           & 2.5  &                                       &      \\ \hline
                           &       &  $\left(L/D\right)_{\mathrm{min}}$           & 0.18 &                                       &      \\ \hline
                           &       &  $N_{\sigma,\mathrm{max}}$                   & 2    &                                       &      \\ \hline
                           &       &  $w_{\mathrm{PT,max}}$                       & 3    &                                       &      \\ \hline
                           &       &  $w_{\mathrm{corr,max}}$                     & 2    &                                       &      \\ \hline
                           &       &  $(b/\sigma_{b})_{\mathrm{cut}}$             & 200  &                                       &      \\ \hline
\end{tabular}
\vspace*{2ex}
\caption{\textit{Values of code parameters used for results presented in this paper.}}
\label{TableValuesOfCodeParameters}
\end{center}
\end{table}
The values of the code parameters that were used to obtain the results
presented in this paper are listed in Table \ref{TableValuesOfCodeParameters}.

\section{Versions of software packages used}
\label{AppendixSoftwareVersions}

\begin{table}[htb]
\begin{center}
\begin{tabular}{||c|c||}
\hline
parameter & version number \\ \hline
Pythia     & 6.4.10    \\
MOKKA      & 06-06-p03 \\
Marlin     & v00-10-03 \\
MarlinReco & v00-10-04 \\
PandoraPFA & v02-02-02 \\
\hline
\end{tabular}
\vspace*{2ex}
\caption{\textit{Version numbers of software packages used in this paper.}}
\label{TableSoftwareVersions}
\end{center}
\end{table}
The version numbers of the various software packages used
to obtain the results presented in this paper are listed in
Table \ref{TableSoftwareVersions}.

\end{document}